\documentclass[aps,prb,
amsmath,amssymb,
 preprint, 
 groupedaddress, superscriptaddress,
 showpacs]{revtex4-2}

\usepackage{bm}
\usepackage{graphicx}
\usepackage{gensymb}
\usepackage{amsmath}
\usepackage{amssymb}
\usepackage{color}
\usepackage{ulem}
\usepackage{hyperref}
\hypersetup{
colorlinks = true,
urlcolor = blue,
linkcolor=blue,
citecolor = blue      
}

\usepackage{xcolor}
\usepackage{xstring}   
\usepackage{etoolbox}

\newcommand{\Add}[1]{\textcolor{black}{#1}}

\def\redkeys{,NohPRL,}

\makeatletter
\let\REVTX@orig@bibitem\bibitem
\newif\ifREVTX@inbib
\REVTX@inbibfalse

\renewcommand{\bibitem}[2][]{%
  \ifREVTX@inbib\endgroup\fi
  \REVTX@orig@bibitem[#1]{#2}%
  \REVTX@inbibtrue
  \begingroup
  \IfSubStr{\redkeys}{,#2,}{\color{black}}{}%
}

\pretocmd{\endthebibliography}{\ifREVTX@inbib\endgroup\REVTX@inbibfalse\fi}{}{}
\makeatother

\begin{document}

\title{Spin-\Add{d}egenerate \Add{b}ulk \Add{b}ands and \Add{t}opological \Add{s}urface \Add{s}tates \Add{a}ssociated with Dirac \Add{n}odal \Add{l}ines in RuO$_2$}%

\author{Takumi Osumi}
\affiliation{Department of Physics, Graduate School of Science, Tohoku University, Sendai 980-8578, Japan}

\author{Kunihiko Yamauchi}
\affiliation{Center for Spintronics Research Network (CSRN), Osaka University, Toyonaka, Osaka 560–8531, Japan}

\author{Seigo Souma}
\thanks{Corresponding authors:\\
seigo.soma.e2@tohoku.ac.jp\\
t-sato@arpes.phys.tohoku.ac.jp}
\affiliation{Advanced Institute for Materials Research (WPI-AIMR), Tohoku University, Sendai 980-8577, Japan}
\affiliation{Center for Science and Innovation in Spintronics (CSIS), Tohoku University, Sendai 980-8577, Japan}

\author{Shubhankar Paul}
\affiliation{Toyota Riken--Kyoto University Research Center (TRiKUC), Kyoto 606-8501, Japan}
\affiliation{Indian Institute of Technology (IIT) Kanpur, Kanpur 208016, India}

\author{Asuka Honma}
\affiliation{Department of Physics, Graduate School of Science, Tohoku University, Sendai 980-8578, Japan}

\author{Kosuke Nakayama}
\affiliation{Department of Physics, Graduate School of Science, Tohoku University, Sendai 980-8578, Japan}



\author{Kenichi Ozawa}
\affiliation{Institute of Materials Structure Science, High Energy Accelerator Research Organization (KEK), Tsukuba, Ibaraki 305-0801, Japan}

\author{Miho Kitamura}
\affiliation{National Institutes for Quantum Science and Technology (QST), Sendai 980-8579, Japan}

\author{Koji Horiba}
\affiliation{National Institutes for Quantum Science and Technology (QST), Sendai 980-8579, Japan}

\author{Hiroshi Kumigashira}
\affiliation{Institute of Multidisciplinary Research for Advanced Materials (IMRAM), Tohoku University, Sendai 980-8577, Japan}

\author{Chiara Bigi}
\affiliation{Synchrotron SOLEIL, L’Orme des Merisiers, Départementale 128, 91190 Saint-Aubin, France}

\author{Fran\c{c}ois Bertran}
\affiliation{Synchrotron SOLEIL, L’Orme des Merisiers, Départementale 128, 91190 Saint-Aubin, France}

\author{Tamio Oguchi}
\affiliation{Center for Spintronics Research Network (CSRN), Osaka University, Toyonaka, Osaka 560–8531, Japan}

\author{Takashi Takahashi}
\affiliation{Department of Physics, Graduate School of Science, Tohoku University, Sendai 980-8578, Japan}

\author{Yoshiteru Maeno}
\affiliation{Toyota Riken--Kyoto University Research Center (TRiKUC), Kyoto 606-8501, Japan}

\author{Takafumi Sato}
\thanks{Corresponding authors:\\
seigo.soma.e2@tohoku.ac.jp\\
t-sato@arpes.phys.tohoku.ac.jp}
\affiliation{Department of Physics, Graduate School of Science, Tohoku University, Sendai 980-8578, Japan}
\affiliation{Advanced Institute for Materials Research (WPI-AIMR), Tohoku University, Sendai 980-8577, Japan}
\affiliation{Center for Science and Innovation in Spintronics (CSIS), Tohoku University, Sendai 980-8577, Japan}
\affiliation{International Center for Synchrotron Radiation Innovation Smart (SRIS), Tohoku University, Sendai 980-8577, Japan}
\affiliation{Mathematical Science Center for Co-creative Society (MathCCS), Tohoku University, Sendai 980-8577, Japan}

\date{\today}

\begin{abstract}
Altermagnets are a novel platform to realize exotic electromagnetic properties distinct from those of conventional ferromagnets and antiferromagnets. We report results of micro-focused angle-resolved photoemission spectroscopy (ARPES) on RuO$_2$, in which its altermagnetic nature has been under fierce debate in connection with crystal-orientation-dependent spintronic functionalities. By elucidating the band structure of the (100), (110) and (101) surfaces of a bulk single crystal using micro-ARPES, we found that, irrespective of the surface orientation, the experimental band structures show a good agreement with the bulk-band calculations for the nonmagnetic phase, but display a severe disagreement with those for the antiferromagnetic phase. Moreover, spin-resolved ARPES signifies a negligible spin polarization in the bulk bands, suggesting the absence of antiferromagnetism and altermagnetic spin splitting. In addition, we identified a nearly flat surface band and a dispersive one near the Fermi level at the (100)/(110) and (101) surfaces, respectively. Our first-principles calculations and analysis of Berry phase attribute these states to the topological surface bands emerging from the bulk Dirac nodal lines around the Fermi level. Our results indicate that such topological surface/interface states must be considered to understand the spintronic functionalities of RuO$_2$, and may provide new insights into its catalytic characteristics.
\end{abstract}


\maketitle

\clearpage
\section{INTRODUCTION}

The coupling of spin and charge is one of key ingredients to give rise to novel quantum phases. When the space inversion symmetry is broken in materials such as noncentrosymmetric bulk crystals and heterostructures, the energy bands become spin split due to the spin-orbit coupling (SOC) \cite{Rashba, Dresselhaus}. These spin-split bands exhibit a momentum-dependent spin texture (spin-momentum locking), leading to novel functionalities in spintronics such as spin-Hall effect and spin-charge conversion \cite{Murakami, SinovaPRL, Kato, SinovaRev}. They are also responsible for various exotic physical properties such as unconventional superconductivity and superconducting diode effect \cite{Sigrist, SYipRev, YanasePRL, LFuPNAS, OnoNat}. Recent theoretical proposals together with the experimental verification of momentum-dependent spin splitting in some collinear antiferromagnets are revolutionizing the field of spintronics and spin-related physics, because the splitting can be realized even without SOC \cite{AhnPRB, NakaNatCom, HayamiJPSJ, SmejkalCHE, YuanPRB, YuanPRM, MaNatCom, SmejkalPRX1, SmejkalPRX2}. In these antiferromagnets called altermagnets \cite{SmejkalPRX1, SmejkalPRX2}, the existence of opposite-spin sublattices connected by a certain symmetry of crystal triggers an anisotropic sign-changing spin splitting associated with the breaking of $PT$ symmetry (combination of space-inversion and time-reversal symmetries). Intriguingly, the magnitude of splitting reaches as large as 1 eV, which is even larger than the SOC-induced spin splitting, making altermagnets an excellent platform to realize novel electromagnetic properties associated with the spin-charge coupling \cite{NakaNatCom, HayamiJPSJ, YuanPRB, MaNatCom, SmejkalPRX1, SmejkalPRX2, YuanPRM}.

Ru$\textrm{O}_2$ with rutile structure [see Fig. 1(a)] has attracted significant attention because of its outstanding properties associated with the spin-charge coupling \cite{AhnPRB, SmejkalCHE, Gonzalez, FengNatEle}, provoking active discussions on whether these properties are attributable to the altermagnetic spin splitting \cite{FengNatEle, Bai2022, KarubePRL, Bose2022, Bai2023, Tschirner, MWangNatCom, GuoAdvSci, WangPRL2024, JeongSHG}. When one assumes a collinear antiferromagnetic (AFM) structure \cite{AhnPRB, Berlijn2017, ZHZhu2019, GregoryRXS}, the oxygen atoms surrounding the Ru atoms break the $PT$ symmetry [Fig. 1(a)], resulting in a $d$-wave-like anisotropic spin splitting of maximally 0.8 eV along the [101] direction with gap nodes in the (100) and (010) planes \cite{SmejkalCHE, FengNatEle}. This anisotropic splitting was theoretically predicted to give rise to a giant anomalous Hall effect and highly efficient spin-charge conversion, both of which show a strong crystallographic-axis dependence \cite{SmejkalCHE, FengNatEle, Gonzalez}. These predictions have been examined experimentally using thin-film samples with various crystal orientations \cite{FengNatEle, Bai2022, KarubePRL, Bose2022, Tschirner, MWangNatCom, GuoAdvSci, WangPRL2024}, as highlighted by the observation of anomalous Hall conductivity comparable to that of Fe in a (110)-oriented sample \cite{FengNatEle} and high spin-charge conversion efficiency in a (101)-oriented sample \cite{Gonzalez, Bose2022}. RuO$_2$ also shows intriguing physical and chemical properties such as anisotropic anomalous Nernst effect \cite{ZhouPRL2024}, anisotropic tunneling magnetoresistance \cite{ShaoNC, ShaoPRL, NohPRL}, terahertz emission \cite{LiuAOM, ZhangTHz}, and catalytic reactivity on the (110) surface \cite{OverScience2000}, while their relationship with the electronic states has yet to be clarified.

While the experimental evidence showing the crystal-orientation-dependence to support the anisotropic spin splitting is accumulating in RuO$_2$ \cite{FengNatEle, KarubePRL, Bai2022, Bai2023, Tschirner, MWangNatCom, GuoAdvSci, WangPRL2024}, there exist some contradictory reports from the neutron scattering and $\mu$SR \cite{HiraishiPRL, Kessler2024} that do not support the magnetic order. Consequently, there remain two key unresolved questions; i.e. (i) whether RuO$_2$ truly exhibits altermagnetism, and (ii) what microscopic electronic structures underlie the observed strong crystal-orientation-dependent transport properties. Although angle-resolved photoemission spectroscopy (ARPES) is considered to be one of essential techniques for addressing these issues through direct observation of the band structure, current ARPES studies are limited to the (110) crystal plane, leaving a comprehensive understanding of its electronic structure incomplete \cite{Rotenberg2018, JovicArxiv2019, JovicACS2021,ARPESSciAdv, LiuARPES,FelserARPES}. Moreover, even within ARPES studies on (110)-oriented single crystals, the results remain highly controversial, with conflicting reports between the paramagnetic (PM) band degeneracy and the altermagnetic band splitting \cite{FelserARPES, LiuARPES} (for chronological overview of the ARPES studies on rutile oxides, see Supplemental Note 1\cite{SM2025}). This controversy cannot be resolved solely by the ARPES studies of the (110) orientation, because key spectral feature to distinguish the altermagnetic band splitting, a flat band near $E_\textrm{F}$ (which will be detailed later), may be attributed either to a spin-split bulk band  \cite{FelserARPES} or to surface states \cite{LiuARPES}. Therefore, it is essential to comprehensively investigate the electronic states of multiple orientations to disentangle surface and bulk states as well as to clarify the universality and dissimilarity in the surface-dependent electronic states.

In this article, we report a spin-resolved ARPES study with a RuO$_2$ bulk single crystal which has three different surface planes of (100), (110) and (101), in combination with first-principles calculations, to address the key questions above. 
\Add{Our ARPES results provide strong spectroscopic evidence supporting the absence of altermagnetic spin splitting in bulk RuO$_2$} irrespective of crystal orientations. We also found energy bands crossing the Fermi level ($E_\textrm{F}$) associated with the surface states (SS) in the ($E$, $k$) region where no corresponding bands exist in the bulk-band calculation. The effective mass of SS strongly depends on the crystal orientation. We discuss implications of the present results in relation to the topological properties.

 \section{Experiments and calculations}
RuO$_2$ single crystals were grown by the vapor-transport method in flowing oxygen \cite{Huang1982}. Polycrystalline RuO$_2$ pellets were placed in an alumina tube and heated at 1250$^{\circ}$C for 168 hours to yield single crystals of the size up to 5 $\times$ 3 $\times$ 2 mm$^3$ at lower temperature locations. Obtained crystals were characterized by x-ray diffraction spectra as well as Laue photographs and electrical resistivity. Temperature ($T$) dependence of the electrical resistivity shows a metallic behavior in the whole $T$ range. Residual resistivity is 0.1 $\Omega$cm for the current along [001], and the residual resistivity ratio (RRR) reaches 400. This value is the highest level among thus-far reported RRR values for the bulk crystal (20–230) \cite{Rogers1969, Ryden1970A, Huang1982, PawulaPRB} and thin films (2–5) \cite{GregoryRXS, FengNatEle, KarubePRL}, confirming the high-quality nature of our single crystal.

 ARPES measurements were performed at BL-28A in Photon Factory (PF) with circularly polarized 40–200 eV photons using a micro beam spot of 12$\times$10 ${\mu}$m$^2$ \cite{KitamuraRSI}. Spin-resolved ARPES measurements were carried out at CASSIOP\'{E}E beamline in SOLEIL with circularly and linear horizontally polarized 46–74 eV photons. Samples were cleaved \textit{in situ} along the (100), (110), and (101) crystal planes in an ultrahigh vacuum of  $\sim{1.0}\times10^{-10}$ Torr, and the successful cleaving was confirmed by looking at the cleaved surface with optical microscope and by checking the symmetry and periodicity of observed band dispersions and Fermi surface (FS). Sample was kept at $T$ = 40 K during ARPES measurements.

First-principles band-structure calculations were carried out by using the projector augmented wave method implemented in the Vienna $\textit{ab initio}$ simulation package (VASP) code \cite{VASP}. The SS was obtained with the surface Green’s function method implemented in WannierTools code \cite{WannierTools} after the maximally localized Wannier functions for Ru-4$d$ and O-2$p$ orbital states were obtained by using Wannier90 code \cite{MLW}. To stabilize the AFM order, we included on-site Coulomb interaction energy $U$ = 2.0 eV for the calculation in the AFM phase, whereas the calculation in the nonmagnetic (NM) phase was obtained without including $U$.

\section{{Electronic structure from (100), (110), and (101) surfaces}}
 Although previous ARPES measurements with bulk single crystals and thin films were carried out on the (110) surface  \cite{Rotenberg2018,  JovicArxiv2019, JovicACS2021,ARPESSciAdv, LiuARPES,FelserARPES}, we found that the bulk crystal can be cleaved nicely also on the (100) and (101) surfaces [indicated by the shaded area in Fig. 1(a)]. These three types of flat surfaces are identified as facets of an as-grown single crystal as shown in Fig. 1(b). We have characterized the orientation of these surfaces by x-ray Laue backscattering measurements [Fig. 1(c)] in which clear and sharp diffraction spots, consistent with the symmetry of each surface, are well recognized.
 
 We first present the band structure and FS with incident photons on the (100) surface which have not been accessed by previous ARPES studies. This plane has a straightforward relationship with the altermagnetic nodal plane. Figure 2(a) shows the calculated bulk FS for the NM (top panel) and AFM (middle and bottom panels) phases. Since the calculation in the PM phase is difficult because of the random alignment of spins, we approximate the experimental PM phase with the NM phase in the calculation. One can immediately recognize several large spin degenerate 3D pockets in the NM phase. On the other hand, the FSs in the AFM phase are spin split due to the altermagnetic band splitting. Because of the $d$-wave nature, the spin-up and spin-down FSs are rotated by $90^{\circ}$ from each other about the $k_z$ axis, and the band splitting vanishes on both the $k_{x\textrm{ or }y} = 0.0$ ($\Gamma$XRZ) and $k_{x\textrm{ or }y} = 1.0$  (XMAR) high-symmetry nodal planes. Away from the high-symmetry planes, such as the $k_x = 0.5$ ($\Delta$YTU) plane, the bands are spin split unless the measured $k$ cut crosses the nodal planes ($k_y$ = 0.0 and 1.0). Such a dramatic difference in the shape of FSs between the NM and AFM phases (hereafter called NM and AFM calculations, respectively) enables identification of plausible magnetic phases by a direct comparison with the ARPES-derived band structures \cite{SmejkalCHE, FengNatEle, AhnPRB, Smolyanyuk}.

Figures 2(b) and 2(c) show the experimental FSs for the (100) surface plotted against the in-plane wave vectors ($k_y$ and $k_z$) obtained for the $k_x \sim 0.0$ plane, which are directly compared with the slice of bulk FSs obtained with the NM and AFM calculations, respectively. The ARPES intensity signifies some characteristic features, such as the absence of spectral weight around the R point and complicated intensity distribution around the $\Gamma$ point. These features have a periodicity of bulk BZ with two-fold symmetry, indicative of the successful cleavage at the (100) surface. As shown in Fig. 2(b), the calculated FSs in the NM phase show an excellent agreement with the ARPES intensity. In particular, rounded triangular pockets (called $\alpha$) around the Z point intersecting along the ZR cut and the absence of FS around the R point seen in the experiment are nicely reproduced in the NM calculation. On the other hand, as shown in Fig. 2(c), the calculated FS in the AFM phase is too simple; it just contains small pockets centered at the $\Gamma$ and Z points, in contradiction to the ARPES result. We have confirmed by systematically changing the on-site coulomb energy $U$ in the AFM calculation that the discrepancy between the experiment and AFM calculation is not attributed to the incorrect choice of $U$ (for details, see \Add{Appendix A}).

To obtain further insights into the band structure, we directly examine the experimental band dispersion. Figures 2(d)--2(i) show the ARPES intensity plotted against the in-plane wave vector ($k_y$ or $k_z$) and binding energy ($E_\textrm{B}$) along high-symmetry $k$ cuts, obtained at selected $k_x$ slices of $k_x$ = 1.0 [Figs. 2(d, e)], 0.5 [Figs. 2(f, g)], and 0.0 [Figs. 2(h, i)]. These plots signify several dispersive bands. One can make a clear comparison between the ARPES results and band calculations by following the band dispersion indicated as B1. At $k_x \sim 0$ [Fig. 2(h)], one can recognize a band at $E_\textrm{B} \sim 0.2 \textrm{ eV}$ at the X point which disperses upward along the XR cut and crosses $E_\textrm{F}$ (green circle). This band disperses back toward higher $E_\textrm{B}$, crossing $E_\textrm{F}$ again (green square) and stays at $E_\textrm{B} \sim 0.5 \textrm{ eV}$ around the R point. Along the RZ cut, it touches $E_\textrm{F}$ around the Z point and rapidly disperses toward higher $E_\textrm{B}$ along the Z$\Gamma$ cut. A direct comparison with the calculated band dispersions in Fig. 2(i) signifies that the NM calculation (black solid curves) semi-quantitatively reproduces the ARPES-derived band dispersion (dark gray open circles); in particular, regarding a shallow hole band topped at slightly above $E_\textrm{F}$ along the XR cut, producing the $\beta$ and $\gamma$ FSs in Fig. 2(b) as well as the hole band topped at slightly above $E_\textrm{F}$ along the RZ$\Gamma$ cut associated with the $\alpha$ pocket in Fig. 2(b). On the other hand, the AFM calculation (red and blue dashed curves; they overlap at $k_x$ = 0) shows a fatal disagreement with the experiment along the XR cut [Fig. 2(i)]. The hole band which crosses $E_\textrm{F}$ in the experiment sinks well below $E_\textrm{F}$ ($E_\textrm{B} \sim 0.2 \textrm{ eV}$) and stays at $\sim 0.9 \textrm{ eV}$ at the R point, away from the experimental one ($\sim 0.5 \textrm{ eV}$). Another hole band at the Z point is topped at far ($\sim 0.2 \textrm{ eV}$) above $E_\textrm{F}$, indicating overestimation of the Fermi wave vector ($k_\textrm{F}$) along the Z$\Gamma$ cut (red arrow). Moreover, the $E_\textrm{F}$ crossing of another band midway between Z and $\Gamma$ in the experiment (green triangle) is not reproduced in the AFM calculation.

We identified an experimental feature which is not reproduced in neither the NM- nor AFM-calculations. That is a nearly flat band in the vicinity of $E_\textrm{F}$ along the $\Gamma$XR and $\Gamma$Z cuts in Fig. 2(h) [near-$E_\textrm{F}$ region of orange circles in Figs. 2(e), 2(g), and 2(i)]. It is assigned to the surface state (SS), as supported by its $h\nu$-independent band dispersion and also by our slab calculations (for details, see \Add{Appendix B and F}). In fact, the $h\nu$ invariance at the $\overline{\Gamma}$ point can be inferred by looking at the commonly identified near-$E_\textrm{F}$ flat band at X, $\Delta$, and $\Gamma$ points in Figs. 2(d), 2(f) and 2(h) all of which are projected onto the $\overline{\Gamma}$ point in the surface Brillouin zone (BZ), despite a strong intensity modulation due to the matrix-element effect of photoelectron intensity. A clear identification of the sharp SS supports good surface quality. Moreover, the crystals used in this study is of substantially higher quality, with RRR of 400, than those in other recent reports in 2019--2025, 10 to 200 \cite{Rogers1969, Ryden1970A, Huang1982, PawulaPRB, GregoryRXS, FengNatEle, KarubePRL}. Previous ARPES reports did not mention the sample quality in terms of RRR as summarized in the table S1 in Supplemental Note 1\cite{SM2025}.

The argument above is further corroborated by a comparison of band dispersions at $k_x \sim 1.0$ between ARPES and calculations in Figs. 2(d) and 2(e). 
Besides the B1 band, the ARPES intensity along the MA cut signifies another band called B2 which shows Dirac-like crossing with the B1 band around $E_\textrm{F}$. This crossing is nicely reproduced by the NM calculation in the similar ($E$, $k$) region [Fig. 2(e)], and is attributed to the bulk Dirac nodal line 1 (DNL1) with the quadruple band degeneracy appearing commonly in rutile oxides \cite{Rotenberg2018, JovicArxiv2019, BinghaiPRB2017, YLChen2019, Nelson2019}. Interestingly, the SS shows the dispersion starting from this DNL1 along the MA cut; we will come back to this point later. Although the AFM calculation also shows a Dirac-crossing behavior at $E_\textrm{F}$ (open red diamond), its $k$ point and the energy dispersion corresponding to the B2 band are very different, again supporting the agreement with the NM calculation. As can be seen in Fig. 2(g), the AFM calculation shows the altermagnetic spin splitting at $k_x$ = 0.5 (the off-nodal plane), leading to the doubling of some bands with respect to the NM case. The ARPES intensity in Fig. 2(f) exhibits no clear signature of such band doubling, as inferred from the smooth evolution of spectral features from $k_x$ = 0 [Fig. 2(h)] to 1.0 [Fig. 2(d)] through $k_x$ = 0.5 [Fig. 2(f)]. We carried out spin-resolved ARPES measurements at $k$ points where the altermagnetic band splitting is theoretically predicted [red line in Fig. 2(f)], but found that the spin polarization for the bulk bands always keeps zero within experimental uncertainty (for details, see \Add{Appendix C}). All these results indicate that the electronic structure of the (100) surface does not support the altermagnetic band splitting.

We found that the ARPES data for the (110) surface [see Fig. 3(a)] share several common features with the (100) surface. The FS mapping against the in-plane wave vectors in Fig. 3(b) obtained in the $k_{\textrm{[110]}} \sim 0$ plane [$\Gamma$ZAM plane; red-shaded plane in Fig. 3(a)] signifies a two-fold-symmetric intensity pattern with the periodicity obeying that of the bulk BZ, consistent with the previous reports \cite{Rotenberg2018, JovicArxiv2019, JovicACS2021, ARPESSciAdv, FelserARPES, LiuARPES}. A side-by-side comparison of the FS mapping with the NM and AFM calculations in Figs. 3(b) and 3(c), respectively, signifies that the overall FS shape surrounding the $\Gamma$ point is reproduced well by the NM calculation whereas the size of $\Gamma$-centered FS in the AFM calculation is much smaller than that of the experiment. 
The ARPES intensity shown in Fig. 3(d) along the $\Gamma$ZAM$\Gamma$ cut signifies several dispersive bands. Comparison of the ARPES-derived band dispersion with the calculated band dispersions for the NM and AFM cases in Fig. 3(e) reveals that the ($E$, $k$) position of experimental bands are semi-quantitatively reproduced in the NM calculation, whereas the AFM calculation apparently shows a mismatch in the energy position. Similarly to the case of the (100) surface, we found evidence for a flat SS around the $\overline{\Gamma}$ point with its surface nature suggested from the $h\nu$-invariance of the peak position as shown in Fig. 3(f). It is remarked here that the previous ARPES study supporting the altermagnetic band splitting \cite{FelserARPES} attributed this flat band to the flat band at 0.4 eV seen in the AFM calculation along the $\Gamma$M cut (red dashed line) by largely shifting the calculated $E_\textrm{F}$ position downward, while this amount of energy shift is unrealistic taking into account of nearly stoichiometric nature of the RuO$_2$ single crystal. It is emphasized here that ARPES measurements on different surface orientations play a crucial role in testing the validity of band assignments, since the surface and bulk band dispersions inherently exhibit different sensitivities to change in surface orientation. Moreover, transport data in RuO$_2$ thin films are known to depend strongly on surface orientation (\textit{e.g.} refs. \cite{SmejkalCHE, SmejkalPRX2, FengNatEle, KarubePRL, Bai2022, Bai2023, Tschirner, MWangNatCom, GuoAdvSci, WangPRL2024,Bose2022,Gonzalez}). Because such transport properties reflect bulk electronic states, clarifying the variations in bulk-band dispersions across surface orientations is highly valuable for identifying the origin of the orientation-dependent transport properties.

Now we turn our attention to the ARPES results for the (101) surface. As shown in Fig. 4(a), the (101) surface is largely tilted from the nodal planes of the altermagnetic spin splitting. Figure 4(b) displays the FS mapping in the $k_{\textrm{[101]}}$$\sim$0 plane, corresponding to the $\Gamma$XVW plane [see right top panel of Fig. 4(a)]. One can recognize a prominent intensity pattern following the symmetry and periodicity of the (101) surface, such as hexagonal and ellipsoidal pockets (dashed curves, called  $\delta$ and $\varepsilon$) centered at the $\Gamma$ point, a small diamond-shaped pocket at the X point ($\kappa$), and X-shaped pattern ($\lambda$) around the V point. A direct comparison of the ARPES intensity with the NM and AFM calculations in Figs. 4(b) and 4(c), respectively, signifies that the calculated FS in the AFM phase simply contains elongated pockets centered at the $\Gamma$ point, showing a disagreement with the ARPES intensity, whereas the calculated FS in the NM phase shows a better agreement with the experiment.

To clarify the band character, we show in Figs. 4(d) and 4(f) the ARPES intensity along representative $k$ cuts in the $k_{\textrm{[101]}}$$\sim$0.0 and 1.0 planes, respectively, compared with the corresponding band dispersions obtained by the NM and AFM calculations in Figs. 4(e) and 4(g). One can recognize in Figs. 4(d) and 4(f) several dispersive bands, some of which cross $E_\textrm{F}$ to contribute to the pockets seen in Figs. 4(b) and 4(c). Although the distinction of more appropriate calculation (NM or AFM) is difficult compared to other surface planes, the experimental broad feature at $E_\textrm{B}$ $\sim$ 0.3--0.6 eV along the RA$\Lambda$R cut in Fig. 4(f) appears to show a better agreement with the NM calculation in Fig. 4(g). Besides these bulk-originated features, one can find in both Figs. 4(d) and 4(f) several dispersive features within $E_\textrm{B} \sim0.3$ eV of $E_\textrm{F}$, while they have no counterparts in the calculation. For example, an inner electron band that produces the hexagonal $\delta$ pocket in Fig. 4(b) appears in the ($E$, $k$) region where the calculated bands in the NM phase are absent. This band displays no band dispersion along the out-of-plane wave vector, $k_{\textrm{[101]}}$ [Figs. 4(d) and 4(f)], supporting its surface origin. One can also recognize a hole band producing a small pocket at the X point associated with the $\kappa$ pocket in Fig. 4(b). This band is not predicted by the NM calculation and has no dispersion along $k_{\textrm{[101]}}$ in the experiment [Fig. 4(g)], and hence, it is also attributed to the SS. Therefore, the (101) surface of RuO$_2$ hosts several dispersive SS that contribute to the formation of surface-derived FS. It is noted here that the inclusion of SOC in the NM calculation slightly improves the matching between the experiment and calculation (see \Add{Appendix D}), but a fatal disagreement is still found in the AFM calculation even when the SOC is included.

 \section{{Topological surface states}}
The present study supports the absence of antiferromagnetism and altermagnetic band splitting in bulk single crystal of RuO$_2$ 
(note that, at the moment, we do not exclude the possibility that a RuO$_2$ thin film shows altermagnetic behavior, because the crystal defects and lattice strain are not the same as those of bulk crystal). 
Besides this conclusion, we propose that the key characteristics of SS observed at the three different crystal orientations are commonly explained in terms of the non-trivial band topology that is responsible for the emergence of topological SS associated with the bulk DNL. This is enabled by the observation of electronic structures for the (100) and (101) orientations that yields two key findings: (i) the identification of several metallic SSs independent of surface orientations, and (ii) strongly orientation-dependent surface band dispersion (flat \textit{vs} dispersive). These results are important for understanding the mechanisms underlying spintronic functionalities, because such functionalities are reported by the thin film experiments \cite{FengNatEle, KarubePRL, Bai2022, Bai2023, Tschirner, MWangNatCom, GuoAdvSci, WangPRL2024,Bose2022,Gonzalez,JeongSHG,LiuAOM,ZhangTHz} and thereby the SSs would play a significant role. Importantly, (i) indicates that the emergence of the SSs could be attributed to the bulk-edge correspondence in the topological band structure like the Dirac-cone SS in a strong topological insulator.

To elaborate on this discussion quantitatively, we examine topological nature of the band structure in terms of Berry phase in topological semimetals \cite{Burkov_nodallline_2011, ChanPRB2016, fang_nodalline_2015, bzdusek_Nat2016, hirayama2017}. Since RuO$_2$ has a rutile structure (space group $P4_2/mnm$) and found to be nonmagnetic from our APRES and first-principles calculations, there exist several nodal lines in the bulk BZ, protected by $PT$ and mirror symmetries ($P$: space-inversion, $T$: time-reversal; see, \Add{Apprndix E}). Among these nodal lines, the DNL located near $E_\textrm{F}$ in the mirror plane, referred to here as DNL1 [Fig. 5(a)] is the most important \Add{DNL} because it is associated with the observed SS. SOC lifts the degeneracy of the DNLs, opening a gap and thereby removing their strict topological protection. However, this does not imply that the original DNLs are topologically trivial. In fact, the Berry curvature generated around the SOC-induced gaps originates from the nontrivial topology of the DNLs that existed before the gap opening, and this curvature can lead to a large intrinsic spin Hall effect \cite{BinghaiPRB2017}. Therefore, to establish the topological nature of RuO$_2$, it is crucial to experimentally verify the existence of surface states that reflect the bulk DNLs, even when the degeneracy is lifted by SOC. Demonstrating such correspondence provides direct evidence that the gapped DNLs inherit their topological character from the underlying band inversion, and thus offers a solid foundation for understanding the spin-current generation mechanism in RuO$_2$. Because the SOC strength in RuO$_2$ is significantly weaker than that in IrO$_2$, this compound provides an ideal platform for investigating the weak-SOC limit where the link between topology and spin transport can be clearly resolved.

 In $PT$-symmetric material without SOC, the topological invariant describing the DNLs is given by the Zak phase ($\theta$):
\begin{eqnarray}
\theta(\bm{k}_\parallel) = -i\sum_{n}^{occ.} \int_{-\pi}^{\pi} {\langle u_n(\bm{k}) | \partial {k_\perp} | u_n(\bm{k})\rangle} dk_\perp
\end{eqnarray}
where $u_n(\bm{k})$ is a Bloch wave function for $n$th occupied band, and $\bm{k}_\parallel$ is the wave vector in the surface BZ (for details, see \Add{Appendix E}) \cite{Vanderbilt_electric_1993,Burkov_nodallline_2011, ChanPRB2016}. In the prototypical nodal-line semimetal, the DNL forms a closed loop (i.e. nodal ring), which gives rise to a well-defined $\theta= \pi$ region inside the projection of the nodal ring on the surface BZ. Due to the bulk-edge correspondence, this $\theta= \pi$ region hosts nearly dispersion-less drumhead-type SSs \cite{Burkov_nodallline_2011, ChanPRB2016, fang_nodalline_2015, bzdusek_Nat2016, hirayama2017}. In contrast to such Dirac nodal ring, the DNL1 extends across the BZ, connecting its opposite boundaries without forming a closed loop. The DNL1 does not enclose any finite area, and therefore no well-defined ``inside" region can be assigned to the  $\theta= \pi$ phase. Nevertheless, we found that the mirror eigenvalues of bulk bands forming the DNL1 are exchanged along (110)/(1$\bar{1}$0) plane, and this change acts as an effective topological boundary within the BZ from which the SS emerges (for details, see \Add{Appendix E}). Analogous to the case of nodal line semimetals, each crossing for DNL1 causes a $\pi$ jump in the Zak phase due to the inversion of the occupied-band mirror eigenvalues. This is confirmed by our calculation of Wilson loop encircling a single DNL1 [see the three red circles in Fig. 5(a)] that consistently yields a value of $\pi$, demonstrating the intrinsic topological nature of DNL1. \Add{We evaluated the Wilson loop around three different $k$-loops to examine the nontrivial nature of DNL1 with particular care, to confirm that the Wilson loop, which represents a local Berry phase, indeed captures the global topology of DNL1.} Since the two mirror planes, (110) and (1$\bar{1}$0), are symmetry-related and induce opposite phase inversions, the two $\pi$ shifts cancel each other, resulting in a net Zak phase of $\theta = 0$ along the [100] directions. \Add{This is manifested in Figs. 5(a) and 5(b) as a blue region in the color map around the projected DNL1, depicted as a wavy cyan line.} Thus, no well-defined $\theta=\pi$ region can be defined, even though SSs can still emerge around the projections of the nodal lines protected by mirror and $PT$ symmetries. Namely, the dispersion of SSs of RuO$_2$ on the (100) surface exhibits a drumhead-like shape but lacks a corresponding topological $\theta= \pi$ character. Consequently, on the (100) and (110) surfaces where DNL1 is doubly projected, two topological SS (assuming spin-degenerate bands) are expected to emerge from the DNL when SOC is neglected, as illustrated in the left panel of Fig. 5(c). Although SOC opens a finite gap along the former DNL, surface-related states remain in the corresponding $k$-region, reflecting the residual band connectivity that exists without SOC, as shown in the right panel of Fig. 5(c).

To make the above qualitative arguments more quantitative, we carried out slab calculations for the (100) surface. As shown in Fig. 5(f), the slab calculation including SOC predicts two prominent SSs along the $\bar{\textrm{X}}\bar{\textrm{S}}$ cut of surface BZ [for surface BZ, see Fig. 5(d)], both appearing within $\pm$0.2 eV of $E_\textrm{F}$. We confirmed that these two SSs have topological nature, because they merge into the DNL1 along the $\bar{\textrm{X}}\bar{\textrm{S}}$ cut of surface BZ, consistent with the schematics in Fig. 5(c) (for details, see \Add{Appendix F}). The corresponding ARPES intensity along the same cut (cut 4) shown in Fig. 5(g) reveals one SS (SS2) with strong spectral weight, while the other SS is not clearly resolved, likely because it is largely unoccupied. Systematic ARPES intensity mapping along a few off-$\bar{\textrm{X}}\bar{\textrm{S}}$ cuts (cuts 1--3) allows us to trace the second SS (SS1). Taken together, our ARPES results demonstrate the presence of SSs connecting the bulk DNL, in good agreement with our first-principles calculations, supporting the topologically nontrivial nature of the DNL and associated SSs. We have also carried out the slab calculations on other surface orientations, (110) and (101), and found that the characteristics of the SS and the DNL is universally explained in terms of nontrivial topology and bulk-edge correspondence (\Add{Appendix F}).

While the present study focuses on the surface of bulk crystal, it would be natural to extend our arguments to the interface between RuO$_2$ (including RuO$_2$ thin film) and other materials. The present study suggests that the outstanding transport properties of RuO$_2$ thin film such as the spin-charge conversion (spin splitting torque)\Add{,  tunneling magnetoresistance,} and other physical properties\Add{\cite{FengNatEle, KarubePRL, Bai2022, Bai2023, Tschirner, MWangNatCom, GuoAdvSci, WangPRL2024,Bose2022,Gonzalez,JeongSHG,LiuAOM,ZhangTHz, NohPRL}. Such properties} involve charge/spin transport across the interface should be discussed with taking into account the existence of prominent topological interface states besides the bulk DNL that strongly contributes to the spin Berry curvature \cite{WangPRL2024, BinghaiPRB2017}.
Our results also show strong crystal-orientation-dependent variation in the band velocity (flat/dispersive) of topological SSs, which would cause significant anisotropy in film properties (for details, see \Add{Appendix G}). In this regard, 
 \Add{an effective model for the surface states and a more elaborate microscopic theoretical framework} should be  \Add{constructed} 
 in future studies. It is also worthwhile to comment on the possibility that the catalytic reaction particularly known for the (110) surface \cite{OverScience2000} may be promoted by the existence of topological SS \cite{BinghaiPRB2017}, because high surface conductivity and robustness of the topological SS would be useful for the catalytic reaction \cite{LiSciAdv2019, LiAngChm2019}, working as a “topological catalyst”. In RuO$_2$, the proximity of the topological SS to $E_\textrm{F}$ [$\sim$ 20 meV ($\sim$ 200 K) of $E_\textrm{F}$; see Fig. 3(f)]  for the (110) and (100) surfaces would promote catalytic reaction at room temperature. Hence,
 we caution that a theoretical model that simply deals with the altermagnetic splitting of bulk bands without incorporating the topological surface/interface states and bulk DNL is insufficient to fully account for some of the experimental results of RuO$_2$.

 \section{Conclusion}
 The present micro-ARPES study \Add{for three different crystal orientations}, in collaboration with first-principles calculations, \Add{provides compelling spectrscopic evidence}  
  \Add{that} 
  bulk RuO$_2$ 
   \Add{does not exhbit }altermagnetic band splitting, \Add{ in agreement with recent experimental reports}. Besides the bulk-band degeneracy, we observed a topological surface states near $E_\textrm{F}$ associated with the bulk Dirac nodal lines irrespective of the crystal orientation. Intriguingly, its characteristics, i.e. the flat vs dispersive nature, was found to be strongly crystal-orientation dependent, and is well captured by our slab calculations. Analysis of topological invariants, Zak phase and Wilson loop for the Dirac nodal lines suggests its non-trivial topology. The present results imply important roles of the topological surface and interface states to account for the exotic physical and chemical properties of RuO$_2$.

\begin{acknowledgments}
We acknowledge G. Mattoni, H. Matsuki, and C. Sow for their technical supports. We also acknowledge L. Balents and J. N. Hausman for useful comments and discussions. This work was supported by JST-CREST (No. JPMJCR18T1), Grant-in-Aid for Scientific Research (JSPS KAKENHI Grant Numbers JP22H01168, JP21H04435 and JP19H01845), Grant-in-Aid for JSPS Research Fellow (No: JP18J20058), and KEK-PF (Proposal number: 2024S2-001). T. Osumi and A.H. thanks GP-Spin and JSPS for financial support. 
\end{acknowledgments}

\clearpage

\appendix
\section{HUBBARD $U$ DEPENDENCE OF THE CALCULATED BAND STRUCTURE}
\label{apxA}To clarify the influence of on-site Coulomb interaction energy (Hubbard $U$) on the calculated Fermi surface (FS) and band structure in the antiferromagnetic (AFM) phase, we compare in Figs. 6(a)--6(c) and 6(d)--6(f) the FS in the two-dimensional wave vector and the ARPES-derived band dispersion, respectively, for the (100) surface of RuO$_2$ at $k_x = 0$ at three different $U$ values of 1.5, 2.0, and 2.5 eV. We found from our calculations that the AFM phase is energetically more stable than the nonmagnetic (NM) phase at $U \geq 2.0$ eV, whereas \textit{vice versa} for $U = 1.5$ eV. Hence, the calculation for $U = 1.5$ eV was carried out for the putative (unstable) AFM phase. One can see in Figs. 6(a)--6(c) that the calculated FS volume at this $k$ slice is systematically reduced upon increasing $U$.

This shrinkage is also seen in the calculated band dispersion in Figs. 6(d)--6(f). Importantly, none of AFM calculations satisfactorily reproduces the experimental data while the matching between the experiment and calculation becomes gradually better upon reducing $U$. This is reasonable because the reduction of $U$ makes the system closer to the NM phase. It is remarked that although the matching between the experiment and calculation becomes the best at $U = 1.5$ eV, the agreement is still poorer than the case of the NM calculation shown in Fig. 2(b) of the main text. Thus, disagreement of the experimental data with the AFM calculation is not due to the incorrect choice of the $U$ value, but due to the intrinsic absence of magnetic order in the RuO$_2$ single crystal.

\section{SURFACE STATE AT THE (100) SURFACE}
\label{apxB}
To clarify the surface or bulk nature of the observed flat band around the $\Gamma$ point for the (100) surface presented in Figs. 2(d)--2(i), we carried out $h\nu$-dependent ARPES measurements. Figure 7(a) shows the $h\nu$-dependence of EDC obtained with a normal emission set-up. One can recognize several features dispersing upon $h\nu$ variation, which originate from the bulk bands. Corresponding to the flat band, a sharp peak is identified in the vicinity of $E_\textrm{F}$ at $h\nu$ = 40 eV. The energy position of this peak is invariant against the $h\nu$ variation, supporting its surface nature, while its intensity is strongly modulated by the photoelectron matrix-element effect. These characteristic spectral features are better seen in the ARPES-intensity plot against $k_x$ and $E_\textrm{B}$ in Fig. 7(b). Comparison with the calculated bands in the NM phase signifies a reasonable correspondence of dispersive features between the experiment and calculation, such as an electron band bottomed at $\sim$ 1.5 eV at the $4^\textrm{th}$ $\Gamma$ point and a holelike band topped at $\sim$ 1.8 eV at the $5^\textrm{th}$ $\Gamma$ point, confirming its bulk nature. On the other hand, there exist no corresponding bands in the calculation that could account for the peak at $E_\textrm{F}$, supporting its surface origin.

\section{SPIN-RESOLVED ARPES DATA}
\label{apxC}
To reconfirm the absence of altermagnetic band splitting in bulk bands, we have carried out spin-resolved ARPES measurements at CASSIOP\'{E}E beamline in synchrotron SOLEIL. Figures 8(a) and 8(c) show the spin-resolved EDCs at $T$ = 40 K measured at a $k$ point midway between the $\Gamma$ and M points for the (100) surface, where the AFM calculation predicts a sizable (more than 0.4 eV) spin splitting. One can immediately recognize no difference between the spin-up and spin-down EDCs along the $y$-axis (we set the $x$ and $y$ axes parallel to the sample surface at the normal emission with the $y$-axis being along [001]). The observed spin polarization is essentially zero within experimental uncertainty as shown in Figs. 8(b) and 8(d). 
Here, we estimate the experimental uncertainty, namely the statistical error of the spin polarization. It is written as ${\Delta}P ( = \sqrt{1/(S^{2}N)})$ where $S$ is  the Sherman function of spin detector and $N$ is the number of detection counts \cite{jozwiak2010}. In our measurements, $S$ was in the range of 0.2--0.3, and the number of counts for each energy point was 60,000--120,000. These values yield an estimated theoretical ${\Delta}P$ of 0.01--0.02, consistent with the experimental standard deviations for the spin polarization shown in Figs. 8(b) and 8(d) [${\Delta}P$ = 0.006--0.016].
We also found that the polarization along the $x$-axis is also zero (not shown). This supports the absence of altermagnetic band splitting at the (100) surface. We have carried out spin-resolved ARPES measurements also for the (110) surface at the $k$ point where the altermagnetic splitting is theoretically predicted in the AFM calculation, but found no clear spin polarization [Figs. 8(e)--8(h)].

\section{INFLUENCE OF SOC TO THE CALCULATED BAND STRUCTURE}
\label{apxD}
To clarify the influence of SOC on the calculated band structure of RuO$_2$, we show in Figs. 9(a) and 9(b) a side-by-side comparison of the calculated band dispersions in the NM phase along the XMARX cut ($k_x = 1.0$) obtained without and with SOC, respectively, overlaid with the corresponding ARPES intensity at the (100) surface. One can see an overall similarity in these calculations, in particular, regarding their ($E$, $k$) position. A closer look reveals a small band doubling in the calculation with SOC, together with a hybridization gap at the intersection in some of these bands, as highlighted by the band crossing point associated with the DNL along the MA cut in Figs. 9(a) and 9(b). This trend is also seen in the calculation for the $\Gamma$XRZ$\Gamma$ line ($k_x = 0.0$) shown in Figs. 9(c) and 9(d). The band doubling predicted in the calculation is so small that it is unlikely to be resolved within our experimental accuracy. At the (101) surface, one can see a similar band doubling in the calculations with SOC [Figs. 9(e)--9(h)], whereas the doubling leads to more complex band dispersions at particular $k$ cuts like the A$\Sigma$ cut due to the band hybridization. The calculation with SOC along the A$\Sigma$ cut shows a relatively good agreement with the ARPES intensity near $E_\textrm{F}$, implying a finite influence of SOC on the experimental spectral feature.

\section{TOPOLOGICAL ANALYSIS OF BULK NODAL LINES}
\label{apxE}
Fermi surfaces in metals possess topological stability when the constituent bands host spin-independent energy degeneracies in the form of nodal points or nodal lines within the BZ \cite{ChiuRMP2016}. The corresponding topological invariant is given by the Berry phase evaluated either on a closed surface enclosing a nodal point or along a line that pierces a closed nodal loop. When such a degeneracy is accompanied by the band inversion, the invariant acquires a nontrivial, quantized value as a “jump” in the Berry phase, and SSs that converge toward the projected degeneracy points emerge. Many compounds exhibit topological nodal structures and associated SS and are therefore classified as topological semimetals.

We evaluate the topology of RuO$_2$ adopting the same framework as these topological semimetals \cite{Burkov_nodallline_2011, fang_nodalline_2015, ChanPRB2016, bzdusek_Nat2016, hirayama2017}. RuO$_2$ has a rutile structure (space group $P4_2/mnm$), and several DNLs exist in the bulk BZ, protected by $PT$ ($P$: space-inversion, $T$: time-reversal) and mirror symmetry. As shown in Fig. 10(a), our bulk band calculations without SOC in the NM phase shown reveal several band crossings in the energy range of approximately $\pm$0.3 eV with respect to $E_\textrm{F}$, originating from band 16 and band 17 (energy bands are sequentially labeled from the lowest one). Along the AM cut, these bands intersect midway between A and M. This Dirac crossing generates an X-shaped nodal line (DNL1) in the 3D BZ at $k_z\sim\pm0.7$ (in units of $\pi/c$), as illustrated by the light-blue curves in Fig. 10(b), which forms the central focus of this study. DNL1 lies on the (110) mirror plane, where bands with opposite mirror eigenvalues cross and keep degeneracy due to the combined mirror and $PT$ symmetry. In addition to DNL1, the crossings of bands 16 and 17 give rise to several other DNLs (DNL2--7) [Figs. 10(a) and 10(b)], but their characteristics are distinct from those of DNL1. For example, as shown in Fig. 10(a), DNL2/3--7 and DNL4 along high-symmetry cuts are located at approximately -0.3 eV and +0.6 eV, respectively, whereas DNL1 is situated much closer to $E_\textrm{F}$ at $\sim-0.064$ eV. While DNL1 forms open nodal lines extending toward neighboring BZs, all the other DNLs exhibit closed-loop structures around the bulk $\Gamma$ point [Fig. 10(b)].

\Add{We evaluate the Zak phase by using equation (1).}
The Zak phase ($\theta$) corresponds to a $Z_2$ invariant, quantized to either 0 or $\pi$ due to the $PT$ symmetry. In the prototypical nodal-line semimetal Ca$_3$P$_2$ (space group $P6_3/mcm$), the DNL forms a closed loop (i.e. nodal ring), which gives rise to a well-defined $\theta = \pi$ region inside the projection of the nodal ring on the surface BZ \cite{ChanPRB2016}. Like RuO$_2$, Ca$_3$P$_2$ possesses mirror symmetry with respect to the (001) plane. The nodal ring appears on the $k_z$ = 0 mirror plane forming a single closed nodal loop near the BZ boundary, in which bands keep degeneracy associated with the combined mirror and $PT$ symmetry. Due to the bulk-edge correspondence, this  $\theta = \pi$ region hosts nearly dispersion-less drumhead-type SSs, whereas no such states appear outside the ring. When a single DNL is projected onto the surface, the Zak phase takes a value of $\pi$  on one side of the projection and 0 on the other, giving rise to topological SS in the region with $\theta = \pi$. In contrast, when two DNLs are projected onto the surface, the Zak phase becomes 2$\pi (= \pi + \pi)$, which is effectively equivalent to 0.

In the case of RuO$_2$, we found that many DNLs with closed loop structure as in Ca$_3$P$_2$ show $\theta = 2\pi$ $(= \pi + \pi)$ which is effectively equivalent to 0 because two DNLs are projected onto the surface. In the right panel of Fig. 10(b), we mapped $\theta$ as a function of the 2D wave vectors in the (100) surface BZ. We found that $\theta$ remains zero (blue region) for the projections of all DNLs except for DNL5, which yields $\theta = \pi$ (red region). This is because all DNLs except DNL5 are elongated along the $k_y$ direction and are therefore doubly projected onto the (100) surface for any $k$. In contrast, DNL5 lies in the $k_x$-$k_z$ plane (i.e. $k_y$ = 0) and is singly projected, resulting in $\theta = \pi$ inside its elongated loop. For the (110) surface [bottom panel of Fig. 10(b)], the dog-bone-shaped nodal loop of DNL6 (yellow curves) is singly projected, giving rise to a region with $\theta = \pi$. In the small area where projections of two DNL6 loops overlap (indicated by six white arrows),  $\theta$ becomes 0. It should be noted, however, that since all of these DNLs lie relatively far from $E_\textrm{F}$, the associated topological SS are buried within the bulk bands and cannot be clearly resolved in our ARPES measurements.

We now turn our attention to DNL1 near $E_\textrm{F}$, which is the primary focus of this study and exhibits topological characteristics distinct from the above case. While the DNL5 forms enclosed loop in the (110) plane (namely $\Gamma$ZAM), the DNL1 extends across the BZ, connecting its opposite boundaries without forming a closed loop. Because the DNL1 is open, it does not enclose any finite area, and therefore no well-defined ``inside" region can be assigned to the $\theta = \pi$ phase. As a result, the Zak phase calculated along the surface-normal [100] direction does not correlate with the presence of the SSs in RuO$_2$. Instead, in the context of mirror symmetry, the mirror eigenvalues of bulk bands are exchanged along (110) plane between the bands 16 and 17. This change in the mirror representation acts as an effective topological boundary within the BZ from which the SS emerges. Thus, the existence of the SS is not governed by the one-dimensional Zak phase like in Ca$_3$P$_2$, but by the band connectivity protected by the mirror and $PT$ symmetries associated with the open nodal lines \cite{BinghaiPRB2017}. In this sense, the SSs on the (100) surface of RuO$_2$ relevant to DNL1 reflect the open nodal-line topology which is not captured by a $\theta = \pi$ phase, in contrast to conventional drumhead SSs defined by a $\theta = \pi$ region.

Analogous to the case of a nodal line protected by mirror and $PT$ symmetries, each crossing for DNL1 causes a $\pi$ jump in the Zak phase due to the inversion of the occupied-band mirror eigenvalues. This is confirmed by our calculation of Wilson loop encircling a single DNL1 [see the three red circles in Fig. 5(a) of the main text] that consistently yields a value of  $\pi$, demonstrating the intrinsic topological nature of DNL1. However, since the two mirror planes, (110) and (1$\bar{1}$0), are symmetry-related and induce opposite phase inversions, the two $\pi$ shifts cancel each other, resulting in a net Zak phase of $\theta$  = 0 along the [100] and [110] directions. Thus, no well-defined $\theta = \pi$ region can be defined, even though SSs can still emerge around the projections of the nodal lines protected by mirror and $PT$ symmetries. Namely, the dispersion of SSs of RuO$_2$ on the (100) surface exhibits a drumhead-like shape but lacks a corresponding topological $\theta = \pi$ character. Consequently, on both the (100) and (110) surfaces where DNL1 is doubly projected, two topological SS (assuming spin-degenerate bands) are expected to emerge from the Dirac node when SOC is neglected, as illustrated in the top panel of Fig. 5(c). Although SOC open a finite gap along the former DNL, surface-related states remain in the corresponding $k$-region, reflecting the residual band connectivity that exists without SOC, as shown in the bottom panel of Fig. 5(c).

\section{SLAB CALCULATIONS}
\label{apxF}
To clarify the characteristics of the observed SSs in more detail, we have carried out slab calculations in the NM phase  for the (100), (110), and (101) surfaces by assuming the oxygen-deficient termination as shown in Figs. 11(a)--(c) [corresponding BZs are also shown in Figs. 11(d)--(f)]. We found that the oxygen-deficient termination reproduces the experimental results relatively well, as \Add{detailed below.} 

\subsection{(100) surface}
Figure 12(a) shows the calculated band dispersion along high-symmetry lines in the surface BZ. In this calculation, SOC was intentionally neglected in order to highlight the connection between the DNL and the topological SS. Two SSs, labeled SS1 and SS2, are connected to DNL1 along the $\bar{\textrm{X}}\bar{\textrm{S}}$ cut, as expected from the schematic in Fig. 5(c). Consistent with the  discussion based on the mirror eigenvalues in previous section (Appendix E), the calculated flat SSs on the (100) surface bridges the projections of nodal lines, instead of filling an enclosed area. Although the DNL1 cannot be characterized by a $\theta = \pi$ phase, the resulting surface bands (SS1 and SS2) exhibit a drumhead-like flat dispersion in certain regions of the BZ (e.g. along $\bar{\Gamma}\bar{\textrm{X}}$) and can be interpreted as a symmetry-protected SSs associated with the nodal-line topology.

Next we have included SOC in our slab calculations. As a result, we obtained a stable solution with a magnetic moment (-0.4$\mu_B$, -0.4$\mu_B$, -0.4$\mu_B$) in the $(x, y, z)$ component induced in the surface oxygen atoms. In Figs. 12(b) and 12(c), we present two cases; one in which the total surface magnetic moment was artificially constrained to zero, and  another in which no such constraint was imposed.  In Fig. 12(b), one can directly identify the influence of SOC. The overall energy dispersion of SS1 and SS2 remain nearly unchanged, except for a very small spin splitting. This indicates that the effect of SOC on the SS dispersion is relatively weak.

In both Fig. 12(a) and 12(b), the flat band appears along the $\bar{\Gamma}\bar{\textrm{X}}$ line, as in the experiment, but is located at $\sim$0.45 eV above $E_\textrm{F}$, in contrast to our ARPES data which show the flat band pinned almost at $E_\textrm{F}$. This discrepancy is resolved in the most stable solution, shown in Fig. 12(c) in which a band-dependent exchange splitting of the SS can be seen; notably, the flat band around the  $\bar{\Gamma}$ point now appears near $E_\textrm{F}$, corresponding to the lower branch of the spin-split SS1 (SS1$\downarrow$). Although the validity of possible surface ferromagnetism remains to be clarified, this slab calculation exhibits much better agreement with the ARPES data, as detailed further below.

\subsection{(110)/(101) Surfaces}
To further validate the topological character of DNL1, we have investigated bulk-edge correspondence in other surface orientations, (110) and (101). In the (110) case, DNL1 extends along a momentum direction perpendicular to the surface [Fig. 10(b)]. This causes DNL1 to be buried within the projection of bulk bands, as illustrated in Fig. 13(b). Specifically, a band crossing point at +0.2 eV is observed midway between $\bar{\Gamma}$ and $\bar{{\textrm{Y}}}$, originating from the crossing point of DNL1 along the ${\Gamma}\textrm{Z}$ line in bulk BZ [Fig. 5(d) in main text]. Similarly, another band crossing point appears near $E_\textrm{F}$, close to the $\bar{{\textrm{Y}}}$ point, originating from the crossing point of DNL1 along the MA line in the bulk BZ. Although the correspondence between the SSs and DNL1 is less clear in the (110) orientation compared to the (100) case, two surface bands can still be identified, extending from the DNL1 crossing point (along ${\Gamma}\textrm{Z}$ line) toward $\bar{\Gamma}$. These bands exhibit downward dispersions and form flat features at $E_\textrm{F}$ and $\sim$-0.2 eV. Importantly, the flat band at $E_\textrm{F}$ shows good correspondence with that observed in the ARPES data on the (110) surface [yellow circles in Fig. 13(c)].

On the other hand, as shown in Fig. 14(a), the (101) plane is tilted with respect to the bulk BZ. As a result, the bulk-band projection area in ($E, k$) space becomes even larger than in the (100) and (110) cases, and the band-gap region around $E_\textrm{F}$ is almost completely suppressed, as shown in Fig. 14(b). Consequently, the SS near $E_\textrm{F}$ is almost fully embedded within the bulk band projection, and the surface component of the band structure (indicated by the size of circles on the dispersion), i.e. the degree of surface localization, is markedly reduced compared to the (110) and (100) surfaces. Due to this strong surface-bulk hybridization and the ill-defined DNL structures in the slab calculation (arising from the aforementioned tilting of the DNLs), it becomes difficult to directly trace the relationship between bulk DNL1 and the SSs on the (101) surface. Nevertheless, as indicated by yellow arrows in Fig. 14(b), one may identify correspondences between the calculated and experimental dispersions [Figs. 14(b) and 14(c)], such as the holelike dispersion along $\bar{\Gamma}\bar{\textrm{Y}}$  ($\Gamma$-X in bulk BZ) seen as the $\delta$ or $\varepsilon$ band in Fig. 14(c), and the electronlike dispersion along $\bar{\textrm{V}}\bar{\textrm{Y}}$ (V-X in the bulk BZ), seen as the $\lambda$ or $\kappa$ band in Fig. 14(c). In addition, the electronlike surface band centered at the $\bar{\textrm{W}}$ point shows agreement between calculation and experiment. Taken together, our comprehensive analysis of the band dispersions obtained from the slab calculations and ARPES across the three surface orientations provides strong evidence for an intimate connection between the bulk DNL1 and the SSs in terms of the nontrivial topology.

\section{WANNIER ORBITALS OF SURFACE STATES}
\label{apxG}

While the location of SS in ($E, k$) space is tightly bound to the projection of the DNLs independent of surface orientation, the surface band velocity exhibits a clear dependence on the surface orientation. This behavior can be explained in terms of the atomic arrangement and orbital character at the surface. As shown in Figs. 15(a) and 15(b), the Wannier orbital of Ru $d_{xy}$, associated with the flat band on the (100) surface, extends along the bonding direction with the O atoms ($b$ axis) at the top surface but not along the $a$ axis. The overlap of wave functions between neighboring atoms is therefore highly anisotropic between the $a$- and $b$-axes, leading to a reduction in band velocity and the flattening of the dispersion along the $a$ axis, consistent with the ARPES observations. On the other hand, on the (101) surface, the Ru orbital exhibits less directionality due to the lower symmetry of this surface [Figs. 15(c) and 15(d)]. This results in metallic-bonding-like characteristics, with large wave-function overlap along both principal axes in the surface plane [i.e. there is no abrupt termination of the surface Wannier orbital, unlike Fig. 15(a)], manifested as a sizable dispersion of the SS along all momentum directions. Thus, although the SSs in RuO$_2$ originate from the topological nature of bulk DNLs, the degree of flatness and metallicity can be effectively tuned by varying the surface orientation.

Here we briefly comment on the robustness of topological SS against chemical modifications. It has been reported that chemical modifications on the RuO$_2$(110) surface can collapse the SS \cite{JovicACS2021}. While this can be explained in terms of the non-topological nature of the SS, generally, it is not well understood to what extent a topological SS can remain robust against chemical modifications. For example, in the prototypical 3D strong topological insulator Bi$_2$Se$_3$, the spectral intensity of the surface band has been reported to almost vanish upon surface adsorption \cite{Biswas2016}. In that case, strong lifetime broadening makes it difficult to judge solely from the spectral weight whether the SS truly disappears. Since RuO$_2$ may share a similar situation, we primarily discuss the topological properties of the SS based on theoretical analyses derived from bulk-edge correspondence, as already detailed in Appendix E and F.

\bibliography{RuO2refs}

@misc{SM2025,
	note = {{See Supplemental Material [url] for a summary of previous ARPES studies on RuO$_2$ and IrO$_2$. The Supplemental Material also contains Refs. \cite{Torun2013,DQHo2025}.}}}

@article{jozwiak2010,
	author = {Jozwiak, C. and Graf, J. and Lebedev, G. and Andresen, N. and Schmid, A. K. and Fedorov, A. V. and El Gabaly, F. and Wan, W. and Lanzara, A. and Hussain, Z.},
	doi = {10.1063/1.3427223},
	issn = {00346748},
	journal = {Review of Scientific Instruments},
	note = {arXiv: 1006.2178},
	number = {5},
	pages = {053904},
	pmid = {20515152},
	title = {A high-efficiency spin-resolved photoemission spectrometer combining time-of-flight spectroscopy with exchange-scattering polarimetry},
	url = {https://pubs.aip.org/aip/rsi/article-abstract/81/5/053904/381748/A-high-efficiency-spin-resolved-photoemission?redirectedFrom=fulltext},
	volume = {81},
	year = {2010}}

@article{ChiuRMP2016,
	author = {Chiu, Ching-Kai and Teo, Jeffrey C. Y. and Schnyder, Andreas P. and Ryu, Shinsei},
	doi = {10.1103/RevModPhys.88.035005},
	issue = {3},
	journal = {Reviews of Modern Physics},
	month = {Aug},
	numpages = {63},
	pages = {035005},
	publisher = {American Physical Society},
	title = {Classification of topological quantum matter with symmetries},
	url = {https://link.aps.org/doi/10.1103/RevModPhys.88.035005},
	volume = {88},
	year = {2016}}

@article{NohPRL,
	author = {Noh, Seunghyeon and Kim, Gye-Hyeon and Lee, Jiyeon and Jung, Hyeonjung and Seo, Uihyeon and So, Gimok and Lee, Jaebyeong and Lee, Seunghyun and Park, Miju and Yang, Seungmin and Oh, Yoon Seok and Jin, Hosub and Sohn, Changhee and Yoo, Jung-Woo},
	doi = {10.1103/nrk5-5zrj},
	issue = {24},
	journal = {Physical Review Letters},
	month = {Jun},
	numpages = {6},
	pages = {246703},
	publisher = {American Physical Society},
	title = {Tunneling Magnetoresistance in Altermagnetic ${\mathrm{RuO}}_{2}$-Based Magnetic Tunnel Junctions},
	url = {https://link.aps.org/doi/10.1103/nrk5-5zrj},
	volume = {134},
	year = {2025}}

@article{Kessler2024,
	author = {Ke{\ss}ler, Philipp and Garcia-Gassull, Laura and Suter, Andreas and Prokscha, Thomas and Salman, Zaher and Khalyavin, Dmitry and Manuel, Pascal and Orlandi, Fabio and Mazin, Igor I. and Valent{\'\i}, Roser and Moser, Simon},
	date = {2024/10/05},
	doi = {10.1038/s44306-024-00055-y},
	id = {Ke{\ss}ler2024},
	isbn = {2948-2119},
	journal = {npj Spintronics},
	number = {1},
	pages = {50},
	title = {{Absence of magnetic order in RuO$_2$: insights from $\mu$SR spectroscopy and neutron diffraction}},
	url = {https://doi.org/10.1038/s44306-024-00055-y},
	volume = {2},
	year = {2024}}

@article{bzdusek_Nat2016,
	author = {Bzdu{\v s}ek, Tom{\'a}{\v s} and Wu, QuanSheng and R{\"u}egg, Andreas and Sigrist, Manfred and Soluyanov, Alexey A.},
	doi = {10.1038/nature19099},
	issn = {0028-0836, 1476-4687},
	journal = {Nature},
	month = oct,
	number = {7623},
	pages = {75--78},
	title = {Nodal-chain metals},
	url = {https://www.nature.com/articles/nature19099},
	urldate = {2025-10-26},
	volume = {538},
	year = {2016}}

@article{fang_nodalline_2015,
	author = {Fang, Chen and Chen, Yige and Kee, Hae-Young and Fu, Liang},
	doi = {10.1103/PhysRevB.92.081201},
	issn = {1098-0121},
	journal = {Physical Review B},
	month = aug,
	number = {8},
	pages = {081201},
	title = {Topological nodal line semimetals with and without spin-orbital coupling},
	url = {https://link.aps.org/doi/10.1103/PhysRevB.92.081201},
	urldate = {2018-06-27},
	volume = {92},
	year = {2015}}

@article{hirayama2017,
	author = {Hirayama, Motoaki and Okugawa, Ryo and Miyake, Takashi and Murakami, Shuichi},
	doi = {10.1038/ncomms14022},
	issn = {2041-1723},
	journal = {Nature Communications},
	month = jan,
	number = {1},
	pages = {14022},
	title = {{Topological Dirac nodal lines and surface charges in fcc alkaline earth metals}},
	url = {http://dx.doi.org/10.1038/ncomms14022},
	volume = {8},
	year = {2017}}

@article{Vanderbilt_electric_1993,
	author = {Vanderbilt, David and King-Smith, R. D.},
	doi = {10.1103/PhysRevB.48.4442},
	journal = {Physical Review B},
	number = {7},
	pages = {4442--4455},
	title = {{Electric polarization as a bulk quantity and its relation to surface charge}},
	url = {https://link.aps.org/doi/10.1103/PhysRevB.48.4442},
	urldate = {2025-10-13},
	volume = {48},
	year = {1993}}

@article{JovicACS2021,
	author = {Jovic, Vedran and Consiglio, Armando and Smith, Kevin E. and Jozwiak, Chris and Bostwick, Aaron and Rotenberg, Eli and Di Sante, Domenico and Moser, Simon},
	doi = {10.1021/acscatal.0c04871},
	journal = {ACS Catalysis},
	month = feb,
	number = {3},
	pages = {1749-1757},
	title = {{Momentum for Catalysis: How Surface Reactions Shape the RuO$_2$ Flat Surface State}},
	url = {https://pubs.acs.org/doi/10.1021/acscatal.0c04871},
	volume = {11},
	year = {2021}}

@unpublished{JovicArxiv2019,
	arxiveprefix = {arXiv},
	arxivid = {1908.02621},
	author = {Jovic, Vedran and Koch, Roland J. and Panda, Swarup K. and Berger, Helmuth and Bugnon, Philippe and Magrez, Arnaud and Thomale, Ronny and Smith, Kevin E. and Biermann, Silke and Jozwiak, Chris and Bostwick, Aaron and Rotenberg, Eli and Sante, Domenico Di and Moser, Simon},
	eprint = {arXiv1908.02621},
	title = {{The Dirac nodal line network in non-symmorphic rutile semimetal RuO$_2$}},
	url = {http://arxiv.org/abs/1908.02621},
	year = {2019}}

@article{Biswas2016,
	author = {Biswas, Deepnarayan and Maiti, Kalobaran},
	doi = {10.1016/j.elspec.2015.11.007},
	journal = {Journal of Electron Spectroscopy and Related Phenomena},
	month = apr,
	pages = {90--94},
	title = {{Exceptional surface states and topological order in Bi$_2$Se$_3$}},
	url = {https://linkinghub.elsevier.com/retrieve/pii/S0368204815002704},
	urldate = {2025-10-13},
	volume = {208},
	year = {2016}}

@article{Burkov_nodallline_2011,
	author = {Burkov, A. A. and Hook, M. D. and Balents, Leon},
	doi = {10.1103/PhysRevB.84.235126},
	journal = {Physical Review B},
	month = dec,
	number = {23},
	pages = {235126},
	title = {{Topological nodal semimetals}},
	url = {https://link.aps.org/doi/10.1103/PhysRevB.84.235126},
	urldate = {2025-10-13},
	volume = {84},
	year = {2011}}

@article{ChanPRB2016,
	author = {Chan, Y.-H. and Chiu, Ching-Kai and Chou, M. Y. and Schnyder, Andreas P.},
	doi = {10.1103/PhysRevB.93.205132},
	journal = {Physical Review B},
	month = may,
	number = {20},
	pages = {205132},
	title = {{Ca$_3$P$_2$ and other topological semimetals with line nodes and drumhead surface states}},
	url = {https://link.aps.org/doi/10.1103/PhysRevB.93.205132},
	urldate = {2025-10-13},
	volume = {93},
	year = {2016}}

@article{LiAngChm2019,
	author = {Li, Guowei and Fu, Chenguang and Shi, Wujun and Jiao, Lin and Wu, Jiquan and Yang, Qun and Saha, Rana and Kamminga, Machteld E. and Srivastava, Abhay K. and Liu, Enke and Yazdani, Aliza N. and Kumar, Nitesh and Zhang, Jian and Blake, Graeme R. and Liu, Xianjie and Fahlman, Mats and Wirth, Steffen and Auffermann, Gudrun and Gooth, Johannes and Parkin, Stuart and Madhavan, Vidya and Feng, Xinliang and Sun, Yan and Felser, Claudia},
	doi = {https://doi.org/10.1002/anie.201906109},
	journal = {Angewandte Chemie International Edition},
	number = {37},
	pages = {13107-13112},
	title = {{Dirac Nodal Arc Semimetal PtSn$_4$: An Ideal Platform for Understanding Surface Properties and Catalysis for Hydrogen Evolution}},
	url = {https://onlinelibrary.wiley.com/doi/abs/10.1002/anie.201906109},
	volume = {58},
	year = {2019}}

@article{LiSciAdv2019,
	author = {Guowei Li and Qiunan Xu and Wujun Shi and Chenguang Fu and Lin Jiao and Machteld E. Kamminga and Mingquan Yu and Harun T{\"u}ys{\"u}z and Nitesh Kumar and Vicky S{\"u}{\ss} and Rana Saha and Abhay K. Srivastava and Steffen Wirth and Gudrun Auffermann and Johannes Gooth and Stuart Parkin and Yan Sun and Enke Liu and Claudia Felser},
	doi = {10.1126/sciadv.aaw9867},
	journal = {Science Advances},
	number = {8},
	pages = {eaaw9867},
	title = {{Surface states in bulk single crystal of topological semimetal Co$_3$Sn$_2$S$_2$ toward water oxidation}},
	url = {https://www.science.org/doi/abs/10.1126/sciadv.aaw9867},
	volume = {5},
	year = {2019}}

@article{Bose2022,
	author = {Arnab Bose and Nathaniel J. Schreiber and Rakshit Jain and Ding-Fu Shao and Hari P. Nair and Jiaxin Sun and Xiyue S. Zhang and David A. Muller and Evgeny Y. Tsymbal and Darrell G. Schlom and Daniel C. Ralph},
	doi = {10.1038/s41928-022-00744-8},
	issn = {2520-1131},
	issue = {5},
	journal = {Nature Electronics},
	month = {5},
	pages = {267-274},
	title = {Tilted spin current generated by the collinear antiferromagnet ruthenium dioxide},
	url = {https://www.nature.com/articles/s41928-022-00744-8},
	volume = {5},
	year = {2022}}

@article{OverScience2000,
	author = {H. Over and Y. D. Kim and A. P. Seitsonen and S. Wendt and E. Lundgren and M. Schmid and P. Varga and A. Morgante and G. Ertl},
	doi = {10.1126/science.287.5457.1474},
	issn = {0036-8075},
	issue = {5457},
	journal = {Science},
	month = {2},
	pages = {1474-1476},
	title = {{Atomic-Scale Structure and Catalytic Reactivity of the RuO$_2$ (110) Surface}},
	url = {https://www.science.org/doi/10.1126/science.287.5457.1474},
	volume = {287},
	year = {2000}}

@article{Nelson2019,
	author = {J. N. Nelson and J. P. Ruf and Y. Lee and C. Zeledon and J. K. Kawasaki and S. Moser and C. Jozwiak and E. Rotenberg and A. Bostwick and D. G. Schlom and K. M. Shen and L. Moreschini},
	doi = {10.1103/PhysRevMaterials.3.064205},
	issn = {2475-9953},
	issue = {6},
	journal = {Physical Review Materials},
	month = {6},
	pages = {064205},
	publisher = {American Physical Society},
	title = {{Dirac nodal lines protected against spin-orbit interaction in IrO$_2$}},
	url = {https://link.aps.org/doi/10.1103/PhysRevMaterials.3.064205},
	volume = {3},
	year = {2019}}

@article{YLChen2019,
	author = {X. Xu and J. Jiang and W. J. Shi and Vicky S{\"u}{\ss} and C. Shekhar and S. C. Sun and Y. J. Chen and S.-K. Mo and C. Felser and B. H. Yan and H. F. Yang and Z. K. Liu and Y. Sun and L. X. Yang and Y. L. Chen},
	doi = {10.1103/PhysRevB.99.195106},
	issn = {2469-9950},
	issue = {19},
	journal = {Physical Review B},
	month = {5},
	pages = {195106},
	publisher = {American Physical Society},
	title = {{Strong spin-orbit coupling and Dirac nodal lines in the three-dimensional electronic structure of metallic rutile IrO$_2$}},
	url = {https://link.aps.org/doi/10.1103/PhysRevB.99.195106},
	volume = {99},
	year = {2019}}

@article{BinghaiPRB2017,
	author = {Yan Sun and Yang Zhang and Chao-Xing Liu and Claudia Felser and Binghai Yan},
	doi = {10.1103/PhysRevB.95.235104},
	issn = {2469-9950},
	issue = {23},
	journal = {Physical Review B},
	month = {6},
	pages = {235104},
	publisher = {American Physical Society},
	title = {{Dirac nodal lines and induced spin Hall effect in metallic rutile oxides}},
	url = {http://link.aps.org/doi/10.1103/PhysRevB.95.235104},
	volume = {95},
	year = {2017}}

@article{ZHZhu2019,
	author = {Z. H. Zhu and J. Strempfer and R. R. Rao and C. A. Occhialini and J. Pelliciari and Y. Choi and T. Kawaguchi and H. You and J. F. Mitchell and Y. Shao-Horn and R. Comin},
	doi = {10.1103/PhysRevLett.122.017202},
	issn = {10797114},
	issue = {1},
	journal = {Physical Review Letters},
	pages = {17202},
	pmid = {31012682},
	publisher = {American Physical Society},
	title = {{Anomalous Antiferromagnetism in Metallic RuO$_2$ Determined by Resonant X-ray Scattering}},
	url = {https://doi.org/10.1103/PhysRevLett.122.017202},
	volume = {122},
	year = {2019}}

@article{Berlijn2017,
	author = {T. Berlijn and P. C. Snijders and O. Delaire and H.-D. Zhou and T. A. Maier and H.-B. Cao and S.-X. Chi and M. Matsuda and Y. Wang and M. R. Koehler and P. R. C. Kent and H. H. Weitering},
	doi = {10.1103/PhysRevLett.118.077201},
	issn = {0031-9007},
	issue = {7},
	journal = {Physical Review Letters},
	month = {2},
	pages = {077201},
	pmid = {28256891},
	title = {{Itinerant Antiferromagnetism in RuO$_2$}},
	url = {https://link.aps.org/doi/10.1103/PhysRevLett.118.077201},
	volume = {118},
	year = {2017}}

@article{LFuPNAS,
	author = {Noah F. Q. Yuan and Liang Fu},
	doi = {10.1073/pnas.2119548119},
	issn = {0027-8424},
	issue = {15},
	journal = {Proceedings of the National Academy of Sciences},
	month = {4},
	pages = {e 2119548119},
	pmid = {35377813},
	title = {{Supercurrent diode effect and finite-momentum superconductors}},
	url = {https://pnas.org/doi/full/10.1073/pnas.2119548119},
	volume = {119},
	year = {2022}}

@article{YanasePRL,
	author = {Akito Daido and Yuhei Ikeda and Youichi Yanase},
	doi = {10.1103/PhysRevLett.128.037001},
	issn = {0031-9007},
	issue = {3},
	journal = {Physical Review Letters},
	month = {1},
	pages = {037001},
	pmid = {35119893},
	publisher = {American Physical Society},
	title = {{Intrinsic Superconducting Diode Effect}},
	url = {https://link.aps.org/doi/10.1103/PhysRevLett.128.037001},
	volume = {128},
	year = {2022}}

@article{SYipRev,
	author = {Yip, Sungkit},
	doi = {https://doi.org/10.1146/annurev-conmatphys-031113-133912},
	issn = {1947-5462},
	journal = {Annual Review of Condensed Matter Physics},
	number = {Volume 5, 2014},
	pages = {15-33},
	publisher = {Annual Reviews},
	title = {{Noncentrosymmetric Superconductors}},
	type = {Journal Article},
	url = {https://www.annualreviews.org/content/journals/10.1146/annurev-conmatphys-031113-133912},
	volume = {5},
	year = {2014}}

@article{SinovaPRL,
	author = {Jairo Sinova and Dimitrie Culcer and Q. Niu and N. A. Sinitsyn and T. Jungwirth and A. H. MacDonald},
	doi = {10.1103/PhysRevLett.92.126603},
	issn = {0031-9007},
	issue = {12},
	journal = {Physical Review Letters},
	month = {3},
	pages = {126603},
	title = {{Universal Intrinsic Spin Hall Effect}},
	url = {https://link.aps.org/doi/10.1103/PhysRevLett.92.126603},
	volume = {92},
	year = {2004}}

@article{Murakami,
	author = {Shuichi Murakami and Naoto Nagaosa and Shou-Cheng Zhang},
	doi = {10.1126/science.1087128},
	issn = {0036-8075},
	issue = {5638},
	journal = {Science},
	month = {9},
	pages = {1348-1351},
	title = {{Dissipationless Quantum Spin Current at Room Temperature}},
	url = {https://www.science.org/doi/10.1126/science.1087128},
	volume = {301},
	year = {2003}}

@article{Sigrist,
	author = {Sigrist, Manfred},
	doi = {10.1063/1.3225489},
	issn = {0094-243X},
	journal = {AIP Conference Proceedings},
	month = {08},
	number = {1},
	pages = {55-96},
	title = {{Introduction to unconventional superconductivity in non‐centrosymmetric metals}},
	url = {https://doi.org/10.1063/1.3225489},
	volume = {1162},
	year = {2009}}

@unpublished{FelserARPES,
	archiveprefix = {arXiv},
	arxivid = {2402.04995},
	author = {Zihan Lin and Dong Chen and Wenlong Lu and Xin Liang and Shiyu Feng and Kohei Yamagami and Jacek Osiecki and Mats Leandersson and Balasubramanian Thiagarajan and Junwei Liu and Claudia Felser and Junzhang Ma},
	eprint = {2402.04995},
	title = {{Observation of Giant Spin Splitting and $d$-wave Spin Texture in Room Temperature Altermagnet RuO$_2$}},
	url = {http://arxiv.org/abs/2402.04995},
	year = {2024}}

@article{GregoryRXS,
	author = {Benjamin Z. Gregory and J{\"o}rg Strempfer and Daniel Weinstock and Jacob P. Ruf and Yifei Sun and Hari Nair and Nathaniel J. Schreiber and Darrell G. Schlom and Kyle M. Shen and Andrej Singer},
	doi = {10.1103/PhysRevB.106.195135},
	issn = {2469-9950},
	issue = {19},
	journal = {Physical Review B},
	month = {11},
	pages = {195135},
	publisher = {American Physical Society},
	title = {{Strain-induced orbital-energy shift in antiferromagnetic RuO$_2$ revealed by resonant elastic x-ray scattering}},
	url = {https://link.aps.org/doi/10.1103/PhysRevB.106.195135},
	volume = {106},
	year = {2022}}

@article{Huang1982,
	author = {Y.S. Huang and H.L. Park and Fred H Pollak},
	doi = {10.1016/0025-5408(82)90166-0},
	issn = {00255408},
	issue = {10},
	journal = {Materials Research Bulletin},
	month = {10},
	pages = {1305-1312},
	title = {{Growth and characterization of RuO$_2$ single crystals}},
	url = {https://linkinghub.elsevier.com/retrieve/pii/0025540882901660},
	volume = {17},
	year = {1982}}

@article{Rogers1969,
	author = {D. B. Rogers and R. D. Shannon and A. W. Sleight and J. L. Gillson},
	doi = {10.1021/ic50074a029},
	issn = {1520510X},
	issue = {4},
	journal = {Inorganic Chemistry},
	pages = {841-849},
	title = {{Crystal chemistry of metal dioxides with rutile-related structures}},
	volume = {8},
	year = {1969}}

@article{Ryden1970A,
	author = {W. D. Ryden and A. W. Lawson and Carl C. Sartain},
	doi = {10.1103/PhysRevB.1.1494},
	issn = {01631829},
	issue = {4},
	journal = {Physical Review B},
	pages = {1494-1500},
	title = {{Electrical transport properties of IrO$_2$ and RuO$_2$}},
	volume = {1},
	year = {1970}}

@article{Smolyanyuk,
	author = {Andriy Smolyanyuk and Igor I. Mazin and Laura Garcia-Gassull and Roser Valent{\'\i}},
	doi = {10.1103/PhysRevB.109.134424},
	issn = {24699969},
	issue = {13},
	journal = {Physical Review B},
	number = {13},
	pages = {134424},
	publisher = {American Physical Society},
	title = {{Fragility of the magnetic order in the prototypical altermagnet RuO$_2$}},
	volume = {109},
	year = {2024}}

@article{ARPESSciAdv,
	author = {Olena Fedchenko and Jan Min{\'a}r and Akashdeep Akashdeep and Sunil Wilfred D'Souza and Dmitry Vasilyev and Olena Tkach and Lukas Odenbreit and Quynh Nguyen and Dmytro Kutnyakhov and Nils Wind and Lukas Wenthaus and Markus Scholz and Kai Rossnagel and Moritz Hoesch and Martin Aeschlimann and Benjamin Stadtm{\"u}ller and Mathias Kl{\"a}ui and Gerd Sch{\"o}nhense and Tomas Jungwirth and Anna Birk Hellenes and Gerhard Jakob and Libor {\v S}mejkal and Jairo Sinova and Hans-Joachim Elmers},
	doi = {10.1126/sciadv.adj4883},
	issn = {2375-2548},
	issue = {5},
	journal = {Science Advances},
	month = {2},
	pages = {eadj4883},
	title = {{Observation of time-reversal symmetry breaking in the band structure of altermagnetic RuO$_2$}},
	url = {http://arxiv.org/abs/2306.02170 https://www.science.org/doi/10.1126/sciadv.adj4883},
	volume = {10},
	year = {2024}}

@article{LiuARPES,
	author = {Jiayu Liu and Jie Zhan and Tongrui Li and Jishan Liu and Shufan Cheng and Yuming Shi and Liwei Deng and Meng Zhang and Chihao Li and Jianyang Ding and Qi Jiang and Mao Ye and Zhengtai Liu and Zhicheng Jiang and Siyu Wang and Qian Li and Yanwu Xie and Yilin Wang and Shan Qiao and Jinsheng Wen and Yan Sun and Dawei Shen},
	doi = {10.1103/PhysRevLett.133.176401},
	issn = {0031-9007},
	issue = {17},
	journal = {Physical Review Letters},
	month = {10},
	pages = {176401},
	title = {{Absence of Altermagnetic Spin Splitting Character in Rutile Oxide RuO$_2$}},
	url = {http://arxiv.org/abs/2409.13504 https://link.aps.org/doi/10.1103/PhysRevLett.133.176401},
	volume = {133},
	year = {2024}}

@article{Rotenberg2018,
	author = {Vedran Jovic and Roland J. Koch and Swarup K. Panda and Helmuth Berger and Philippe Bugnon and Arnaud Magrez and Kevin E. Smith and Silke Biermann and Chris Jozwiak and Aaron Bostwick and Eli Rotenberg and Simon Moser},
	doi = {10.1103/PhysRevB.98.241101},
	issn = {24699969},
	issue = {24},
	journal = {Physical Review B},
	pages = {241101(R)},
	publisher = {American Physical Society},
	title = {{Dirac nodal lines and flat-band surface state in the functional oxide RuO$_2$}},
	volume = {98},
	year = {2018}}

@article{VASP,
	author = {G Kresse and J Furthm{\"u}ller},
	doi = {10.1103/PhysRevB.54.11169},
	issn = {0163-1829},
	issue = {16},
	journal = {Physical Review B},
	month = {10},
	pages = {11169-11186},
	publisher = {American Physical Society},
	title = {{Efficient iterative schemes for ab initio total-energy calculations using a plane-wave basis set}},
	url = {https://link.aps.org/doi/10.1103/PhysRevB.54.11169},
	volume = {54},
	year = {1996}}

@article{GuoAdvSci,
	author = {Yaqin Guo and Jing Zhang and Zengtai Zhu and Yuan yuan Jiang and Longxing Jiang and Chuangwen Wu and Jing Dong and Xing Xu and Wenqing He and Bin He and Zhiheng Huang and Luojun Du and Guangyu Zhang and Kehui Wu and Xiufeng Han and Ding-fu Shao and Guoqiang Yu and Hao Wu},
	doi = {10.1002/advs.202400967},
	issn = {21983844},
	issue = {25},
	journal = {Advanced Science},
	pages = {2400967},
	pmid = {38626379},
	title = {{Direct and Inverse Spin Splitting Effects in Altermagnetic RuO$_2$}},
	volume = {11},
	year = {2024}}

@article{HiraishiPRL,
	author = {M Hiraishi and H Okabe and A Koda and R Kadono and T Muroi and D Hirai and Z Hiroi},
	doi = {10.1103/PhysRevLett.132.166702},
	issn = {1079-7114},
	issue = {16},
	journal = {Physical Review Letters},
	pages = {166702},
	publisher = {American Physical Society},
	title = {{Nonmagnetic Ground State in RuO$_2$ Revealed by Muon Spin Rotation}},
	url = {https://doi.org/10.1103/PhysRevLett.132.166702},
	volume = {132},
	year = {2024}}

@unpublished{JeongSHG,
	arxiveprefix = {arXiv:},
	arxivid = {arXiv:2405.05838},
	author = {Seung Gyo Jeong and In Hyeok Choi and Sreejith Nair and Luca Buiarelli and Bita Pourbahari and Jin Young Oh and Nabil Bassim and Ambrose Seo and Woo Seok Choi and Rafael M. Fernandes and Turan Birol and Liuyan Zhao and Jong Seok Lee and Bharat Jalan},
	eprint = {arXiv:2405.05838},
	title = {{Altermagnetic Polar Metallic phase in Ultra-Thin Epitaxially-Strained RuO$_2$ Films}},
	url = {http://arxiv.org/abs/2405.05838},
	year = {2024}}

@article{LiuAOM,
	author = {Yu Liu and Hua Bai and Yuna Song and Zhihao Ji and Shitao Lou and Zongzhi Zhang and Cheng Song and Qingyuan Jin},
	doi = {10.1002/adom.202300177},
	issn = {21951071},
	issue = {16},
	journal = {Advanced Optical Materials},
	pages = {2300177},
	title = {{Inverse Altermagnetic Spin Splitting Effect-Induced Terahertz Emission in RuO$_2$}},
	volume = {11},
	year = {2023}}

@article{PawulaPRB,
	author = {Florent Pawula and Ali Fakih and Ramzy Daou and Sylvie H{\'e}bert and Natalia Mordvinova and Oleg Lebedev and Denis Pelloquin and Antoine Maignan},
	doi = {10.1103/PhysRevB.110.064432},
	issn = {24699969},
	issue = {6},
	journal = {Physical Review B},
	month = {8},
	pages = {064432},
	publisher = {American Physical Society},
	title = {{Multiband transport in RuO$_2$}},
	url = {https://link.aps.org/doi/10.1103/PhysRevB.110.064432},
	volume = {110},
	year = {2024}}

@article{ShaoNC,
	author = {Ding Fu Shao and Shu Hui Zhang and Ming Li and Chang Beom Eom and Evgeny Y. Tsymbal},
	doi = {10.1038/s41467-021-26915-3},
	issn = {20411723},
	issue = {1},
	journal = {Nature Communications},
	pages = {7061},
	pmid = {34862380},
	publisher = {Springer US},
	title = {{Spin-neutral currents for spintronics}},
	volume = {12},
	year = {2021}}

@article{ShaoPRL,
	author = {Ding Fu Shao and Yuan Yuan Jiang and Jun Ding and Shu Hui Zhang and Zi An Wang and Rui Chun Xiao and Gautam Gurung and W. J. Lu and Y. P. Sun and Evgeny Y. Tsymbal},
	doi = {10.1103/PhysRevLett.130.216702},
	issn = {10797114},
	issue = {21},
	journal = {Physical Review Letters},
	pages = {216702},
	pmid = {37295086},
	title = {{N\'{e}el Spin Currents in Antiferromagnets}},
	volume = {130},
	year = {2023}}

@article{WangPRL2024,
	author = {Z. Q. Wang and Z. Q. Li and L. Sun and Z. Y. Zhang and K. He and H. Niu and J. Cheng and M. Yang and X. Yang and G. Chen and Z. Yuan and H. F. Ding and B. F. Miao},
	doi = {10.1103/PhysRevLett.133.046701},
	issn = {0031-9007},
	issue = {4},
	journal = {Physical Review Letters},
	month = {7},
	pages = {046701},
	title = {{Inverse Spin Hall Effect Dominated Spin-Charge Conversion in (101) and (110)-Oriented RuO$_2$ Films}},
	url = {https://link.aps.org/doi/10.1103/PhysRevLett.133.046701},
	volume = {133},
	year = {2024}}

@article{ZhangTHz,
	author = {Sheng Zhang and Yongwei Cui and Shunjia Wang and Haoran Chen and Yaxin Liu and Wentao Qin and Tongyang Guan and Chuanshan Tian and Zhe Yuan and Lei Zhou and Yizheng Wu and Zhensheng Tao},
	doi = {10.1117/1.AP.5.5.056006},
	issn = {25775421},
	issue = {5},
	journal = {Advanced Photonics},
	pages = {056006},
	title = {{Nonrelativistic and nonmagnetic terahertz-wave generation via ultrafast current control in anisotropic conductive heterostructures}},
	volume = {5},
	year = {2023}}

@article{ZhouPRL2024,
	author = {Xiaodong Zhou and Wanxiang Feng and Run-Wu Zhang and Libor {\v S}mejkal and Jairo Sinova and Yuriy Mokrousov and Yugui Yao},
	doi = {10.1103/PhysRevLett.132.056701},
	issn = {0031-9007},
	issue = {5},
	journal = {Physical Review Letters},
	month = {1},
	pages = {056701},
	title = {{Crystal Thermal Transport in Altermagnetic RuO$_2$}},
	url = {http://arxiv.org/abs/2305.01410 https://doi.org/10.1103/PhysRevLett.132.056701 https://link.aps.org/doi/10.1103/PhysRevLett.132.056701},
	volume = {132},
	year = {2024}}

@article{Gonzalez,
	author = {Rafael Gonz{\'a}lez-Hern{\'a}ndez and Libor {\v S}mejkal and Karel V{\'y}born{\'y} and Yuta Yahagi and Jairo Sinova and Tom{\'a}{\v s} Jungwirth and Jakub {\v Z}elezn{\'y}},
	doi = {10.1103/PhysRevLett.126.127701},
	issn = {0031-9007},
	issue = {12},
	journal = {Physical Review Letters},
	month = {3},
	pages = {127701},
	pmid = {33834809},
	title = {{Efficient Electrical Spin Splitter Based on Nonrelativistic Collinear Antiferromagnetism}},
	url = {https://link.aps.org/doi/10.1103/PhysRevLett.126.127701},
	volume = {126},
	year = {2021}}

@article{KitamuraRSI,
	author = {Miho Kitamura and Seigo Souma and Asuka Honma and Daisuke Wakabayashi and Hirokazu Tanaka and Akio Toyoshima and Kenta Amemiya and Tappei Kawakami and Katsuaki Sugawara and Kosuke Nakayama and Kohei Yoshimatsu and Hiroshi Kumigashira and Takafumi Sato and Koji Horiba},
	doi = {10.1063/5.0074393},
	issn = {0034-6748},
	issue = {3},
	journal = {Review of Scientific Instruments},
	month = {3},
	pages = {033906},
	pmid = {35364976},
	publisher = {AIP Publishing, LLC},
	title = {{Development of a versatile micro-focused angle-resolved photoemission spectroscopy system with Kirkpatrick--Baez mirror optics}},
	url = {https://doi.org/10.1063/5.0074393 https://aip.scitation.org/doi/10.1063/5.0074393},
	volume = {93},
	year = {2022}}

@article{AhnPRB,
	author = {Kyo Hoon Ahn and Atsushi Hariki and Kwan Woo Lee and Jan Kune{\v s}},
	doi = {10.1103/PhysRevB.99.184432},
	issn = {24699969},
	issue = {18},
	journal = {Physical Review B},
	pages = {184432},
	publisher = {American Physical Society},
	title = {{Antiferromagnetism in RuO$_2$ as $d$-wave Pomeranchuk instability}},
	volume = {99},
	year = {2019}}

@article{Bai2022,
	author = {H. Bai and L. Han and X. Y. Feng and Y. J. Zhou and R. X. Su and Q. Wang and L. Y. Liao and W. X. Zhu and X. Z. Chen and F. Pan and X. L. Fan and C. Song},
	doi = {10.1103/PhysRevLett.128.197202},
	issn = {0031-9007},
	issue = {19},
	journal = {Physical Review Letters},
	month = {5},
	pages = {197202},
	pmid = {35622053},
	title = {{Observation of Spin Splitting Torque in a Collinear Antiferromagnet RuO$_2$}},
	url = {https://link.aps.org/doi/10.1103/PhysRevLett.128.197202},
	volume = {128},
	year = {2022}}

@article{Bai2023,
	author = {H. Bai and Y. C. Zhang and Y. J. Zhou and P. Chen and C. H. Wan and L. Han and W. X. Zhu and S. X. Liang and Y. C. Su and X. F. Han and F. Pan and C. Song},
	doi = {10.1103/PhysRevLett.130.216701},
	issn = {0031-9007},
	issue = {21},
	journal = {Physical Review Letters},
	month = {5},
	pages = {216701},
	pmid = {37295074},
	title = {{Efficient Spin-to-Charge Conversion via Altermagnetic Spin Splitting Effect in Antiferromagnet RuO$_2$}},
	url = {https://link.aps.org/doi/10.1103/PhysRevLett.130.216701},
	volume = {130},
	year = {2023}}

@article{FengNatEle,
	author = {Zexin Feng and Xiaorong Zhou and Libor {\v S}mejkal and Lei Wu and Zengwei Zhu and Huixin Guo and Rafael Gonz{\'a}lez-Hern{\'a}ndez and Xiaoning Wang and Han Yan and Peixin Qin and Xin Zhang and Haojiang Wu and Hongyu Chen and Ziang Meng and Li Liu and Zhengcai Xia and Jairo Sinova and Tom{\'a}{\v s} Jungwirth and Zhiqi Liu},
	doi = {10.1038/s41928-022-00866-z},
	issn = {2520-1131},
	issue = {11},
	journal = {Nature Electronics},
	month = {11},
	pages = {735-743},
	title = {{An anomalous Hall effect in altermagnetic ruthenium dioxide}},
	url = {http://arxiv.org/abs/2002.08712 https://www.nature.com/articles/s41928-022-00866-z},
	volume = {5},
	year = {2022}}

@article{KarubePRL,
	author = {Shutaro Karube and Takahiro Tanaka and Daichi Sugawara and Naohiro Kadoguchi and Makoto Kohda and Junsaku Nitta},
	doi = {10.1103/PhysRevLett.129.137201},
	issn = {10797114},
	issue = {13},
	journal = {Physical Review Letters},
	pages = {137201},
	pmid = {36206408},
	publisher = {American Physical Society},
	title = {{Observation of Spin-Splitter Torque in Collinear Antiferromagnetic RuO$_2$}},
	url = {https://doi.org/10.1103/PhysRevLett.129.137201},
	volume = {129},
	year = {2022}}

@article{MWangNatCom,
	author = {Meng Wang and Katsuhiro Tanaka and Shiro Sakai and Ziqian Wang and Ke Deng and Yingjie Lyu and Cong Li and Di Tian and Shengchun Shen and Naoki Ogawa and Naoya Kanazawa and Pu Yu and Ryotaro Arita and Fumitaka Kagawa},
	doi = {10.1038/s41467-023-43962-0},
	issn = {20411723},
	issue = {1},
	journal = {Nature Communications},
	pages = {8240},
	pmid = {38086819},
	title = {{Emergent zero-field anomalous Hall effect in a reconstructed rutile antiferromagnetic metal}},
	url = {https://www.nature.com/articles/s41467-023-43962-0},
	volume = {14},
	year = {2023}}

@article{SmejkalCHE,
	author = {Libor {\v S}mejkal and Rafael Gonz{\'a}lez-Hern{\'a}ndez and T. Jungwirth and J. Sinova},
	doi = {10.1126/sciadv.aaz8809},
	issn = {2375-2548},
	issue = {23},
	journal = {Science Advances},
	month = {6},
	pages = {eaaz8809},
	pmid = {32548264},
	title = {{Crystal time-reversal symmetry breaking and spontaneous Hall effect in collinear antiferromagnets}},
	url = {https://www.science.org/doi/10.1126/sciadv.aaz8809},
	volume = {6},
	year = {2020}}

@article{Tschirner,
	author = {Teresa Tschirner and Philipp Ke{\ss}ler and Ruben Dario Gonzalez Betancourt and Tommy Kotte and Dominik Kriegner and Bernd B{\"u}chner and Joseph Dufouleur and Martin Kamp and Vedran Jovic and Libor Smejkal and Jairo Sinova and Ralph Claessen and Tomas Jungwirth and Simon Moser and Helena Reichlova and Louis Veyrat},
	doi = {10.1063/5.0160335},
	issn = {2166532X},
	issue = {10},
	journal = {APL Materials},
	pages = {101103},
	publisher = {AIP Publishing, LLC},
	title = {{Saturation of the anomalous Hall effect at high magnetic fields in altermagnetic RuO$_2$}},
	url = {https://doi.org/10.1063/5.0160335},
	volume = {11},
	year = {2023}}

@article{MLW,
	author = {Arash A. Mostofi and Jonathan R. Yates and Young-Su Lee and Ivo Souza and David Vanderbilt and Nicola Marzari},
	doi = {10.1016/j.cpc.2007.11.016},
	issn = {00104655},
	issue = {9},
	journal = {Computer Physics Communications},
	month = {5},
	pages = {685-699},
	title = {{wannier90: A tool for obtaining maximally-localised Wannier functions}},
	url = {https://linkinghub.elsevier.com/retrieve/pii/S0010465507004936},
	volume = {178},
	year = {2008}}

@article{WannierTools,
	author = {QuanSheng Wu and ShengNan Zhang and Hai-Feng Song and Matthias Troyer and Alexey A. Soluyanov},
	doi = {10.1016/j.cpc.2017.09.033},
	issn = {00104655},
	journal = {Computer Physics Communications},
	month = {3},
	pages = {405-416},
	title = {{WannierTools: An open-source software package for novel topological materials}},
	url = {https://linkinghub.elsevier.com/retrieve/pii/S0010465517303442},
	volume = {224},
	year = {2018}}

@article{HayamiJPSJ,
	author = {Satoru Hayami and Yuki Yanagi and Hiroaki Kusunose},
	doi = {10.7566/JPSJ.88.123702},
	issn = {0031-9015},
	issue = {12},
	journal = {Journal of the Physical Society of Japan},
	month = {12},
	pages = {123702},
	title = {{Momentum-Dependent Spin Splitting by Collinear Antiferromagnetic Ordering}},
	url = {https://journals.jps.jp/doi/10.7566/JPSJ.88.123702},
	volume = {88},
	year = {2019}}

@article{NakaNatCom,
	author = {Makoto Naka and Satoru Hayami and Hiroaki Kusunose and Yuki Yanagi and Yukitoshi Motome and Hitoshi Seo},
	doi = {10.1038/s41467-019-12229-y},
	issn = {2041-1723},
	issue = {1},
	journal = {Nature Communications},
	month = {9},
	pages = {4305},
	pmid = {31541112},
	publisher = {Springer US},
	title = {{Spin current generation in organic antiferromagnets}},
	url = {http://dx.doi.org/10.1038/s41467-019-12229-y https://www.nature.com/articles/s41467-019-12229-y},
	volume = {10},
	year = {2019}}

@article{YuanPRB,
	author = {Lin-Ding Yuan and Zhi Wang and Jun-Wei Luo and Emmanuel I. Rashba and Alex Zunger},
	doi = {10.1103/PhysRevB.102.014422},
	issn = {2469-9950},
	issue = {1},
	journal = {Physical Review B},
	month = {7},
	pages = {014422},
	title = {{Giant momentum-dependent spin splitting in centrosymmetric low-Z antiferromagnets}},
	url = {https://link.aps.org/doi/10.1103/PhysRevB.102.014422 http://arxiv.org/abs/1912.12689 http://dx.doi.org/10.1103/PhysRevB.102.014422},
	volume = {102},
	year = {2020}}

@article{MaNatCom,
	author = {Hai-Yang Ma and Mengli Hu and Nana Li and Jianpeng Liu and Wang Yao and Jin-Feng Jia and Junwei Liu},
	doi = {10.1038/s41467-021-23127-7},
	issn = {2041-1723},
	issue = {1},
	journal = {Nature Communications},
	month = {5},
	pages = {2846},
	pmid = {33990597},
	publisher = {Springer US},
	title = {{Multifunctional antiferromagnetic materials with giant piezomagnetism and noncollinear spin current}},
	url = {http://dx.doi.org/10.1038/s41467-021-23127-7 https://www.nature.com/articles/s41467-021-23127-7},
	volume = {12},
	year = {2021}}

@article{SmejkalPRX2,
	author = {Libor {\v S}mejkal and Jairo Sinova and Tomas Jungwirth},
	doi = {10.1103/PhysRevX.12.040501},
	issn = {2160-3308},
	issue = {4},
	journal = {Physical Review X},
	month = {12},
	pages = {040501},
	title = {{Emerging Research Landscape of Altermagnetism}},
	url = {https://link.aps.org/doi/10.1103/PhysRevX.12.040501},
	volume = {12},
	year = {2022}}

@article{SmejkalPRX1,
	author = {Libor {\v S}mejkal and Jairo Sinova and Tomas Jungwirth},
	doi = {10.1103/PhysRevX.12.031042},
	issn = {2160-3308},
	issue = {3},
	journal = {Physical Review X},
	month = {9},
	pages = {031042},
	title = {{Beyond Conventional Ferromagnetism and Antiferromagnetism: A Phase with Nonrelativistic Spin and Crystal Rotation Symmetry}},
	url = {https://link.aps.org/doi/10.1103/PhysRevX.12.031042},
	volume = {12},
	year = {2022}}

@article{YuanPRM,
	author = {Lin-Ding Yuan and Zhi Wang and Jun-Wei Luo and Alex Zunger},
	doi = {10.1103/PhysRevMaterials.5.014409},
	issn = {2475-9953},
	issue = {1},
	journal = {Physical Review Materials},
	month = {1},
	pages = {014409},
	publisher = {American Physical Society},
	title = {{Prediction of low-Z collinear and noncollinear antiferromagnetic compounds having momentum-dependent spin splitting even without spin-orbit coupling}},
	url = {https://doi.org/10.1103/PhysRevMaterials.5.014409 https://link.aps.org/doi/10.1103/PhysRevMaterials.5.014409},
	volume = {5},
	year = {2021}}

@article{OnoNat,
	author = {Fuyuki Ando and Yuta Miyasaka and Tian Li and Jun Ishizuka and Tomonori Arakawa and Yoichi Shiota and Takahiro Moriyama and Youichi Yanase and Teruo Ono},
	doi = {10.1038/s41586-020-2590-4},
	isbn = {4158602025},
	issn = {14764687},
	issue = {7821},
	journal = {Nature},
	pages = {373-376},
	pmid = {32814888},
	publisher = {Springer US},
	title = {{Observation of superconducting diode effect}},
	url = {http://dx.doi.org/10.1038/s41586-020-2590-4},
	volume = {584},
	year = {2020}}

@article{Kato,
	author = {Y. K. Kato and R. C. Myers and A. C. Gossard and D. D. Awschalom},
	doi = {10.1126/science.1105514},
	issn = {00368075},
	issue = {5703},
	journal = {Science},
	pages = {1910-1913},
	title = {{Observation of the spin hall effect in semiconductors}},
	volume = {306},
	year = {2004}}

@article{SinovaRev,
	author = {Jairo Sinova and Sergio O. Valenzuela and J. Wunderlich and C. H. Back and T. Jungwirth},
	doi = {10.1103/RevModPhys.87.1213},
	issn = {15390756},
	issue = {4},
	journal = {Reviews of Modern Physics},
	pages = {1213-1260},
	title = {{Spin Hall effects}},
	volume = {87},
	year = {2015}}

@article{Dresselhaus,
	author = {G Dresselhaus},
	doi = {10.1103/PhysRev.100.580},
	issue = {2},
	journal = {Physical Review},
	month = {10},
	pages = {580-586},
	publisher = {American Physical Society},
	title = {{Spin-Orbit Coupling Effects in Zinc Blende Structures}},
	url = {https://link.aps.org/doi/10.1103/PhysRev.100.580},
	volume = {100},
	year = {1955}}

@article{Rashba,
	author = {Yu. A Bychkov and E I Rashba},
	issue = {2},
	journal = {P. Zh. Eksp. Teor. Fiz.},
	pages = {66},
	title = {{Properties of a 2D electron gas with lifted spectral degeneracy}},
	url = {http://jetpletters.ru/ps/0/article_19121.shtml},
	volume = {39},
	year = {1984}}

\clearpage
\begin{figure}
\begin{center}
\includegraphics[width=4in]{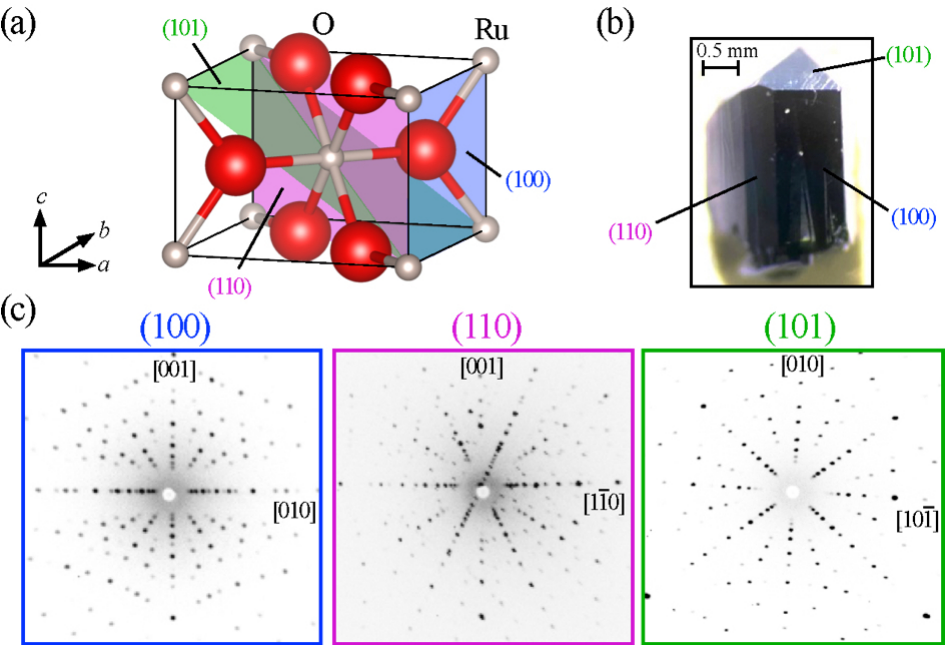}
\hspace{0.2in}
\caption{\textbf{X-ray characterization of RuO$_2$ crystal on different surface planes.} (a), (b) Crystal structure and photograph of a typical RuO$_2$ single crystal, respectively. (c) X-ray Laue images for (100), (110), and (101) surfaces [shaded areas in (a)].
}
\end{center}
\end{figure}

\clearpage
\begin{figure*}
\begin{center}
\includegraphics[width=6.5in]{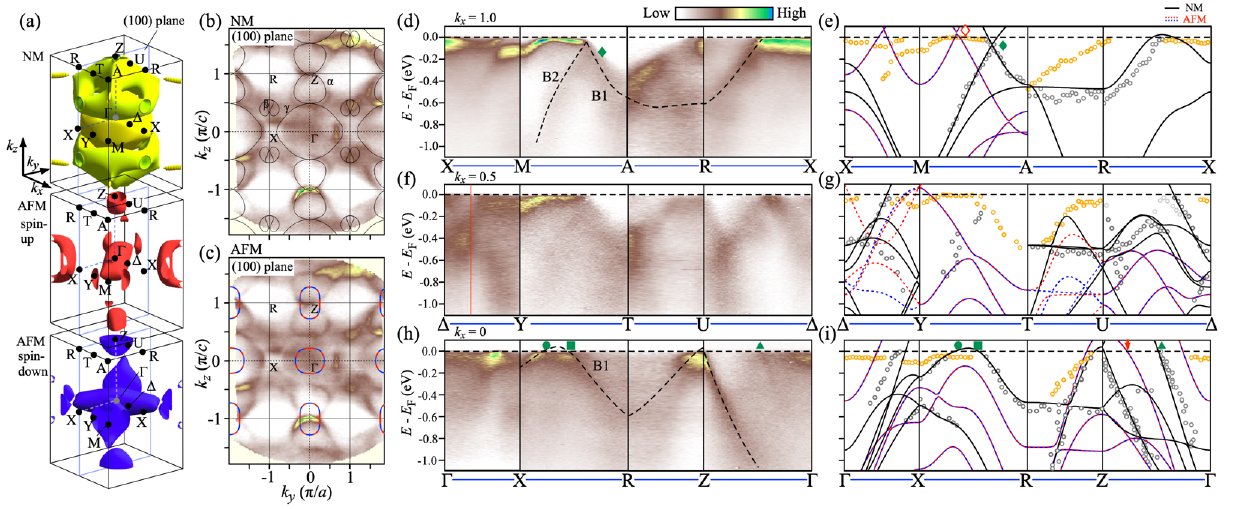}
\caption{\textbf{ARPES spectra obtained from the (100) surface of RuO$_2$ crystal.} (a) Calculated FSs obtained by (top) NM calculation, and AFM calculation for (middle) spin up and (bottom) spin-down states. In the calculation for the AFM phase, we neglected the SOC. (b), (c) FS mapping at $T$ = 40 K for the (100) surface at $k_x = 0.0$ obtained with $h\nu$ = 110 eV, overlaid with the FS obtained by NM and AFM calculations, respectively. Spin-up and spin-down FSs in (c) indicated by red/blue dashed curves completely overlap in this plane. (d), (e) ARPES-intensity plot obtained along the principal $k$ cut (XMARX cut) at $k_x = 1.0$ (in the unit of $\pi/a$) and corresponding calculated band dispersions. Black solid curves are for the NM phase, and red/blue dashed curves (they overlap along this $k$ cut) are for the AFM phase. Dashed curves in (d) are guides to the eye to trace the experimental band dispersion. Experimental band dispersions extracted by tracing the peak position in energy distribution curves (EDCs) are indicated by open circles in (e). (f, g), (h, i) Same as (d), (e) but obtained at $k_x$ = 0.5 and 0.0, respectively. Light gray circles in (e), (g), and (i) highlight features which may originate from the spectral broadening along the wave vector perpendicular to the (100) plane. Orange circles highlight possible SS.
Note that a band within 0.4 eV of $E_\textrm{F}$  along the AR, TU, and RZ cuts (orange circles), which has no counterpart in the bulk band calculation in the NM phase, is attributed to the SS, because it shows a good correspondence with the surface band dispersion obtained by the slab calculations along the $\bar{\textrm{S}}\bar{\textrm{Y}}$ cut shown in Fig. 5(f).
}
\end{center}
\end{figure*}

\clearpage
\begin{figure*}
\begin{center}
\includegraphics[width=6.5in]{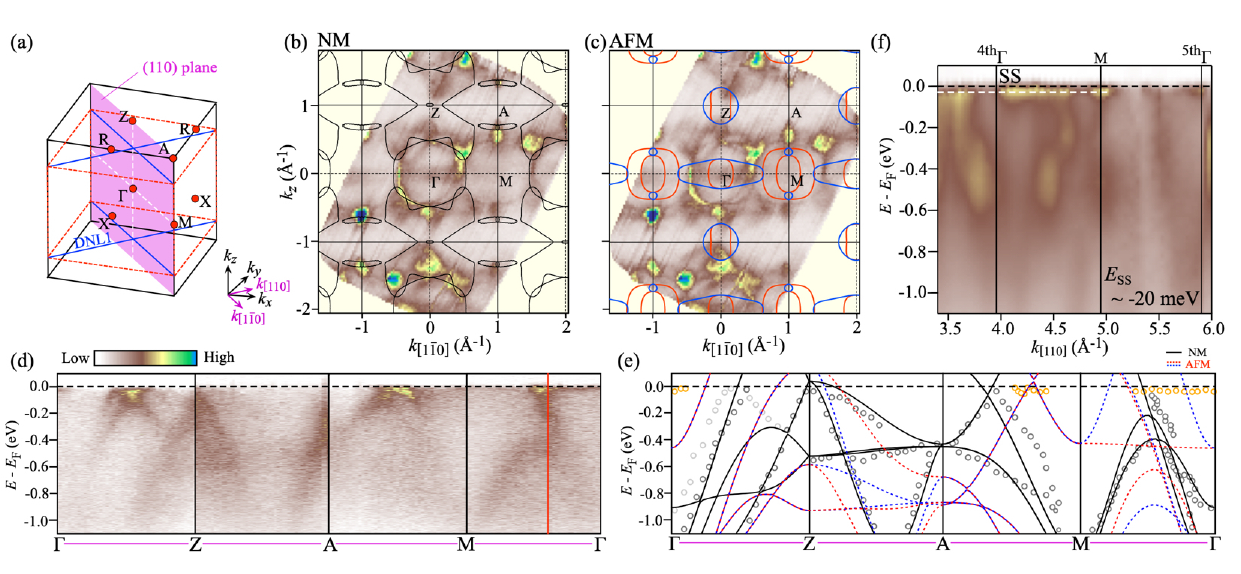}
\caption{\textbf{ARPES spectra obtained from the (110) surface of RuO$_2$ crystal.} (a) Bulk BZ and $k$ plane ($k_{\textrm{[110]}} = 0.0$) where the FS mapping shown in (b) and (c) was obtained. Blue lines represent DNL predicted by the DFT calculation \cite{BinghaiPRB2017}. For simplicity, DNL are depicted as straight lines. Region enclosed by the DNL, i.e. bulk-band-inverted $k$ region, is indicated by red dashed box. (b), (c) FS mapping at the (110) surface obtained at $k_{\textrm{[110]}} = 0.0$ with $h\nu$ = 126 eV, overlaid with the calculated FS obtained by NM (b) and AFM (c) calculations. (d) ARPES intensity plotted against $E_\textrm{B}$ and wave vector along the $\Gamma$ZAM$\Gamma$ cut. (e) ARPES-derived band dispersion (open circles) overlaid with the calculated band dispersions obtained by the NM calculation (black curves) and AFM calculation (red and blue dashed curves). The ARPES results are explained well with the NM calculation, but sharply disagree with the AFM calculation. (f) Normal-emission ARPES intensity plotted against $E_\textrm{B}$ and wave vector along the $k_{\textrm{[110]}}$ axis. SS refers to the surface states.
}
\end{center}
\end{figure*}

\clearpage
\begin{figure}
\begin{center}
\includegraphics[width=3.7in]{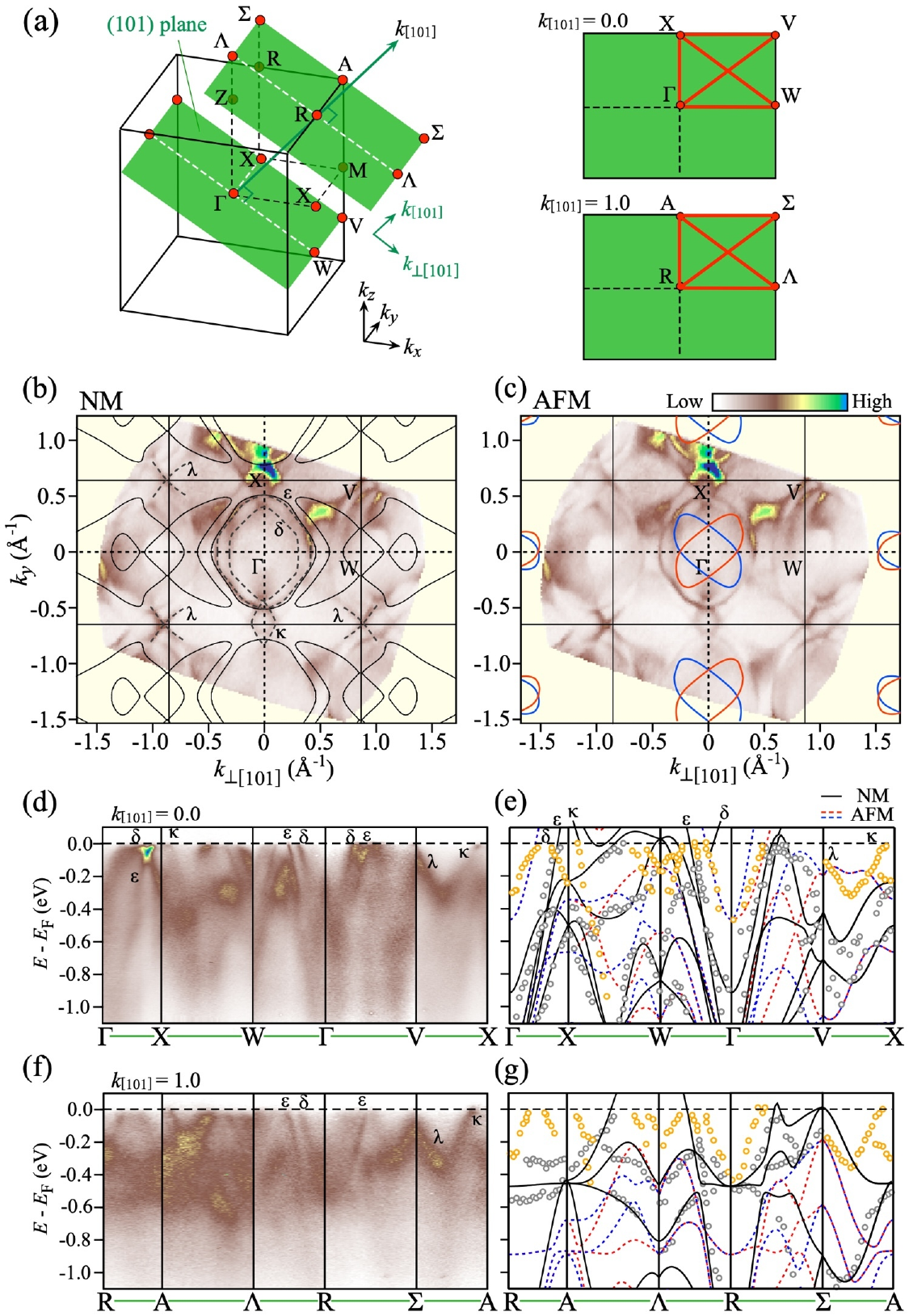}
\caption{\textbf{ARPES spectra obtained from the (101) surface of RuO$_2$ crystal.} (a) Bulk BZ and $k$ planes ($k_{\textrm{[101]}} = 0.0$ and 1.0) in which the FS mapping and ARPES intensity shown in (b)--(g) were obtained. Right panels indicate the $k$ cuts (red solid lines) where the ARPES-intensity plots in (d) and (f) are obtained. Note that V, W, $\Sigma$, and $\Lambda$ points are not the high-symmetry points. (b), (c) FS mapping at the (101) surface at $k_{\textrm{[101]}} = 0.0$ at $h\nu$ = 83 eV, overlaid with the FS obtained by the NM and AFM calculations, respectively. Red and blue curves in (c) represent spin-up and spin-down FSs. Note that the calculated FS in the NM phase shows a better agreement with the experiment, in particular, regarding the presence of $\varepsilon$ pocket and the existence of FS around the W point. A careful look further reveals that the experimental $\delta$, $\kappa$, and $\lambda$ FSs show no counterparts in the calculation, suggesting their SS origin. (d), (e) ARPES-intensity plot obtained along the principle $k$ cut ($\Gamma$XW$\Gamma$VX cut) at $k_{\textrm{[101]}} = 0.0$ and corresponding calculated band dispersions in the NM (black solid curves) and AFM (red and blue dashed curves) phases, respectively. Experimental band dispersions are shown by open circles in (e). (f), (g) Same as (d) and (e) but measured along the RA$\Lambda$R$\Sigma$A cut at $k_{\textrm{[101]}} = 1.0$.
}
\end{center}
\end{figure}

\clearpage
\begin{figure}
\begin{center}
\includegraphics[width=5.7in]{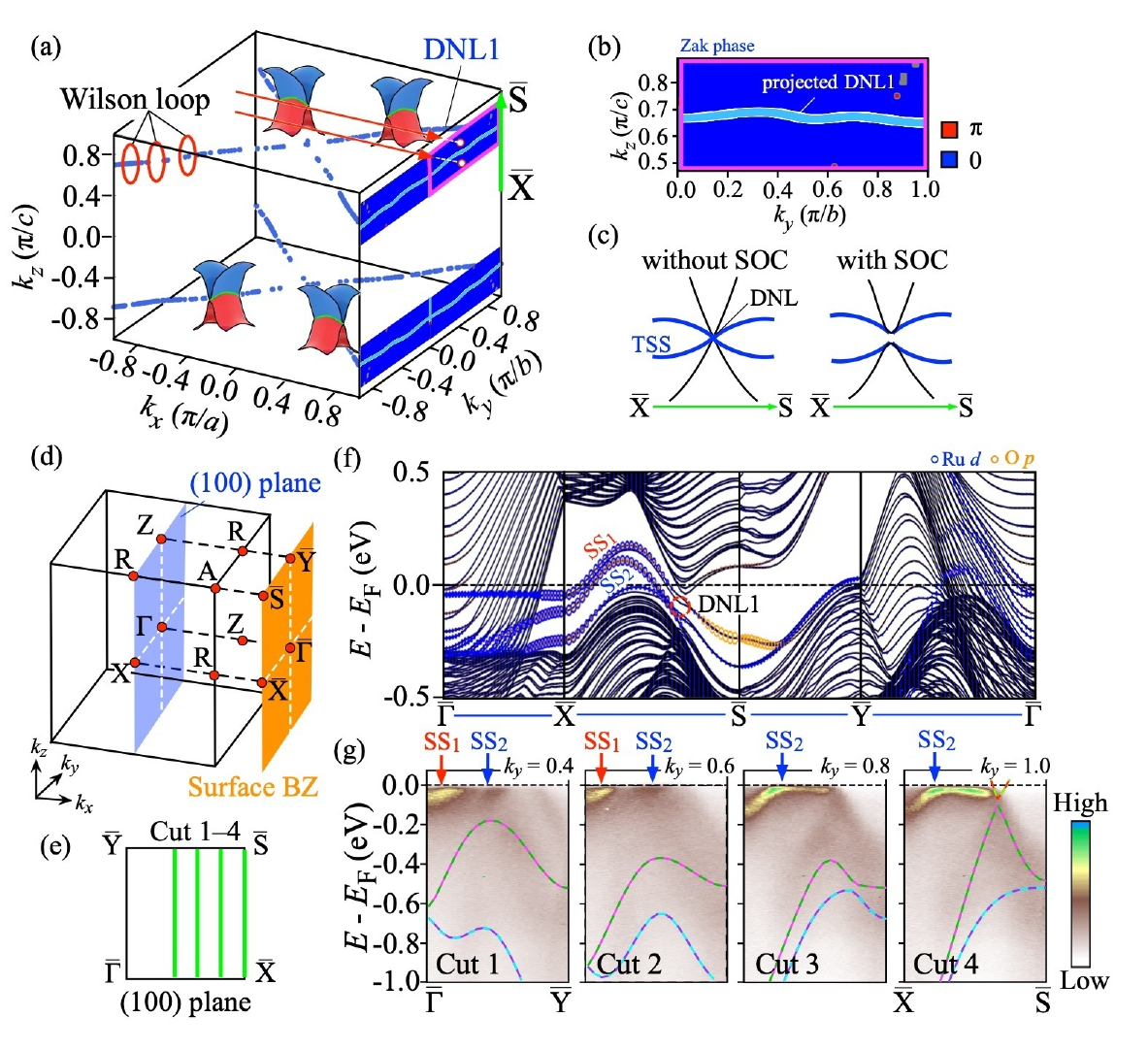}
\caption{\textbf{Topological nature of surface band on (100) surface.} (a) DNL1 (light-blue dots) in bulk BZ. The Wilson loop associated with DNL1 was calculated along the $k$-loops indicated by red circles. A map of Zak phase on the (100) plane is shown in the right side. \Add{The four red-blue 3D maps represent the schematic illustration of energy band dispersion for DNL1 with the Dirac point extending one-dimensionally in $k$-space. The blue and red color indicate upper and lower branches, respectively.} (b) Enlarged view of the Zak phase mapping around DNL1 \Add{shown in magenda line in (a)}. \Add {Light blue wavy line is  a projection of the DNL1 onto the (100) surface.} (c) Schematic illustration of the energy dispersion of the bulk DNL (black curve) and the associated topological SS (blue curve) along a $k$ cut (here the $\bar{\textrm{X}}\bar{\textrm{S}}$ cut) crossing the DNL, shown without (left) and with (right) SOC. (d) Bulk BZ and (100) surface BZ. (e) Representative $k$ cuts (cuts 1-4) in which ARPES intensity in (g) were obtained. (f) Slab calculation including SOC along high-symmetry lines in the surface BZ. Circle size indicates the surface spectral weight, while the Ru 4$d$ and O 2$p$ orbital characters are shown in blue and yellow, respectively. (g) ARPES intensity plots along four representative $k$ cuts, indicated by green lines in (e).
}
\end{center}
\end{figure}

\begin{figure}[h]
\includegraphics[width=4.5 in]{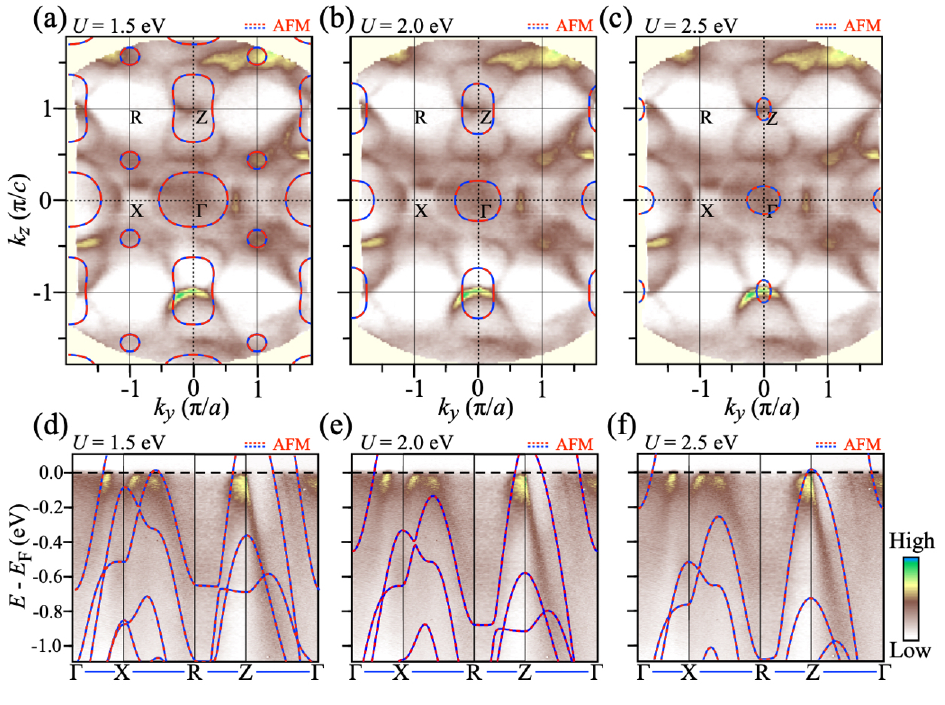}
\caption{\textbf{Comparison of Fermi surface between experiment and AFM calculations with different $U$ values in RuO$_2$.} (a)-(c) FS mapping of RuO$_2$ from ARPES at $T$ = 40 K for the (100) surface at $k_x = 0.0$ obtained with $h\nu$ = 110 eV, overlaid with the calculated FS obtained by the AFM calculation without SOC at $U$ = 1.5, 2.0, and 2.5 eV, respectively. Spin-down and spin-up FSs are indicated by red and blue dashed curves, respectively. (d)--(f) ARPES-intensity mapping for the (100) surface along the $\Gamma$XRZ$\Gamma$ cut at $k_x = 0.0$, compared with the AFM calculation with $U$ = 1.5, 2.0, and 2.5 eV, respectively. For $U$ = 1.5 eV, the AFM state presented in (a) and (d) actually has a higher energy than the NM state.
}
\end{figure}

\begin{figure}[h]
\includegraphics[width=4.5 in]{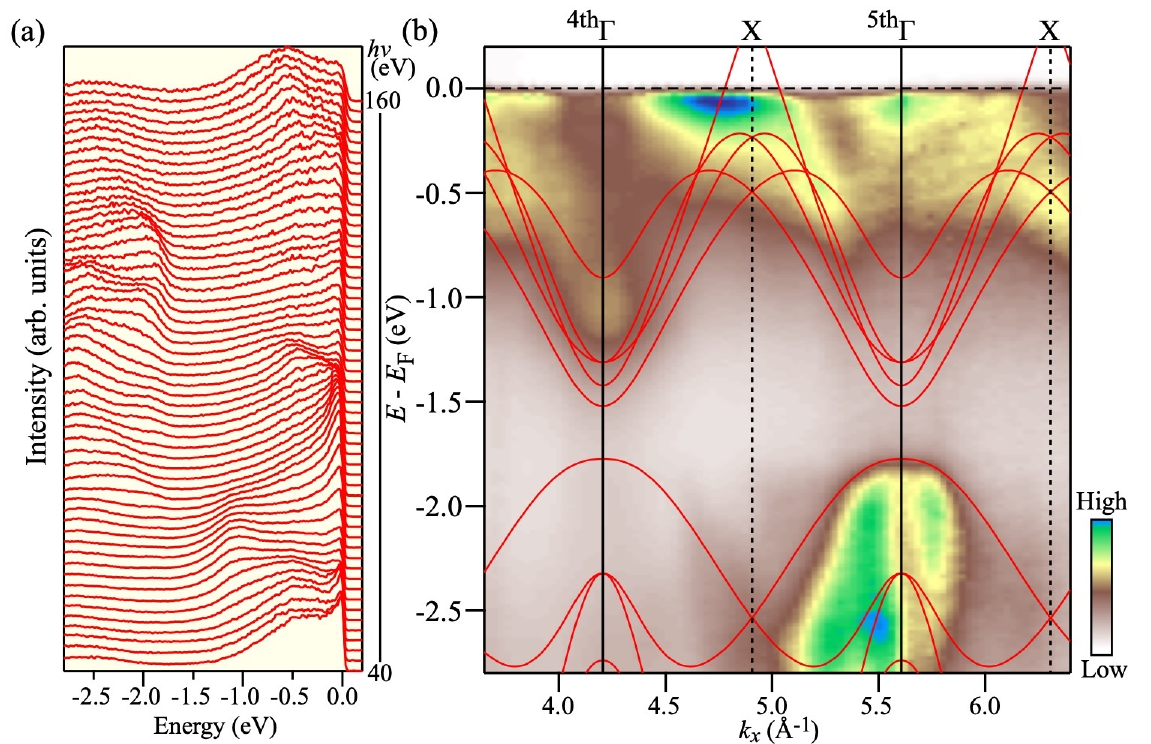}
\caption{ \textbf{Photon-energy-dependent ARPES data in RuO$_2$.} (a) $h\nu$ dependence of EDC obtained at the normal-emission setup for the (100) surface of RuO$_2$. (b) Corresponding ARPES intensity in the $h\nu$ range of 40--160 eV plotted against $k_x$ and $E_\textrm{B}$, together with the calculated band dispersions in the NM phase (red curves) along $k_x$ axis corresponding to the $\Gamma$X cut. We estimated the inner potential to be $V_0 = 15$ eV from the periodicity of the observed band dispersion.
}
\end{figure}

\clearpage

\begin{figure}[h]
\includegraphics[width=4.0 in]{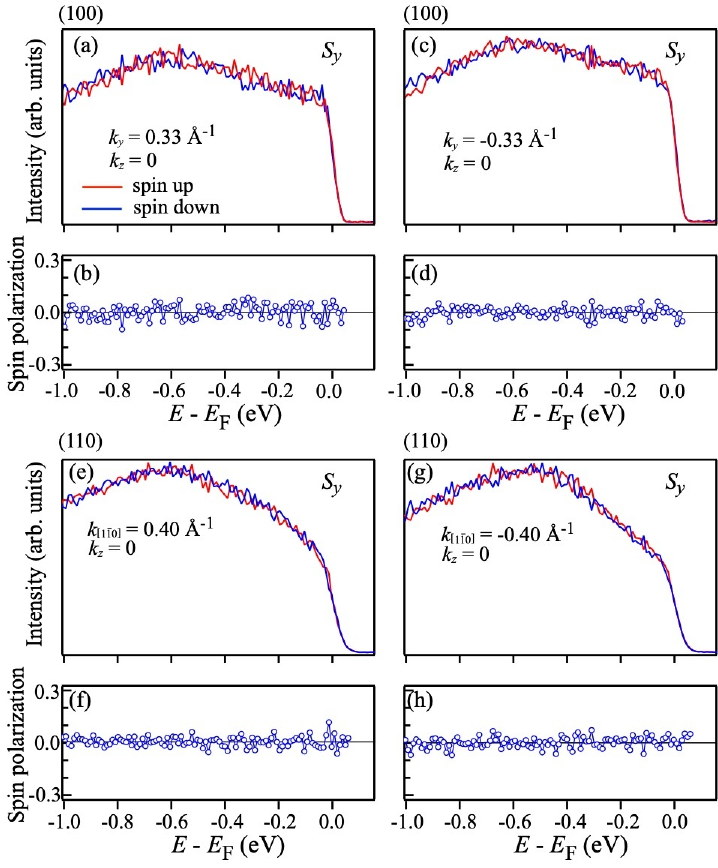}
\caption{\textbf{Spin-resolved EDCs and spin polarizations in RuO$_2$.} (a)-(d) Spin-resolved EDCs and $E_\textrm{B}$ dependence of spin polarization at the wave vectors $k_y = \pm 0.33$ \AA$^{-1}$ along the $\Delta$Y cut [see red line in Fig. 2(f) in the main text] at the (100) surface. (e)--(h) Same as (a)--(d) but for the (110) surface, obtained at the wave vectors $k_{\textrm{[1\={1}0]}} = \pm 0.40$ \AA$^{-1}$ along the $\Gamma$M cut [see red line in Fig. 3(d) in the main text]. There is no spin splitting with the experimental precision, supporting the absence of magnetic order in RuO$_2$.
}
\end{figure}

\clearpage

\begin{figure}
\includegraphics[width=4 in]{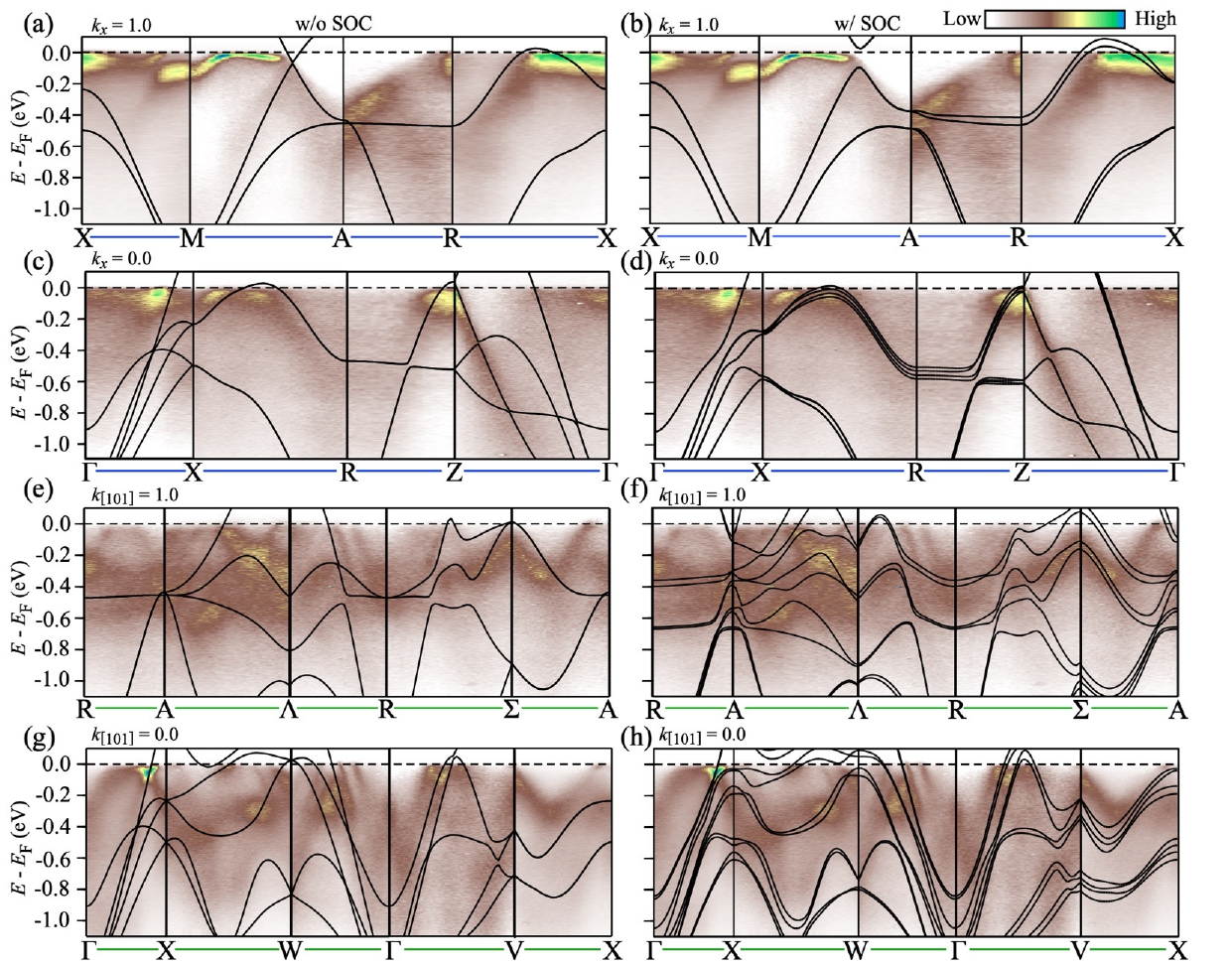}
\caption{ \textbf{Comparison of band dispersion between experiment and NM calculations with and without SOC.} (a), (b) Calculated band dispersions in the NM phase of RuO$_2$ without and with SOC, respectively, along the principal $k$ cut (XMARX cut) at $k_x = 1.0$ for the (100) surface. ARPES intensity [same as Fig. 2(d) in the main text] is also overlaid. (c), (d) Same as (a) and (b), respectively, but along the $\Gamma$XRZ$\Gamma$ cut at $k_x = 0.0$ for the (100) surface. (e), (f) Same as (a) and (b), respectively, but along the principle $k$ cut at $k_\textrm{[101]}= 1.0$ for the (101) surface. (g), (h) Same as (e) and (f), respectively, but at $k_\textrm{[101]}= 0.0$ for the (101) surface.
}
\end{figure}

\begin{figure}[h]
\includegraphics[width=6.5 in]{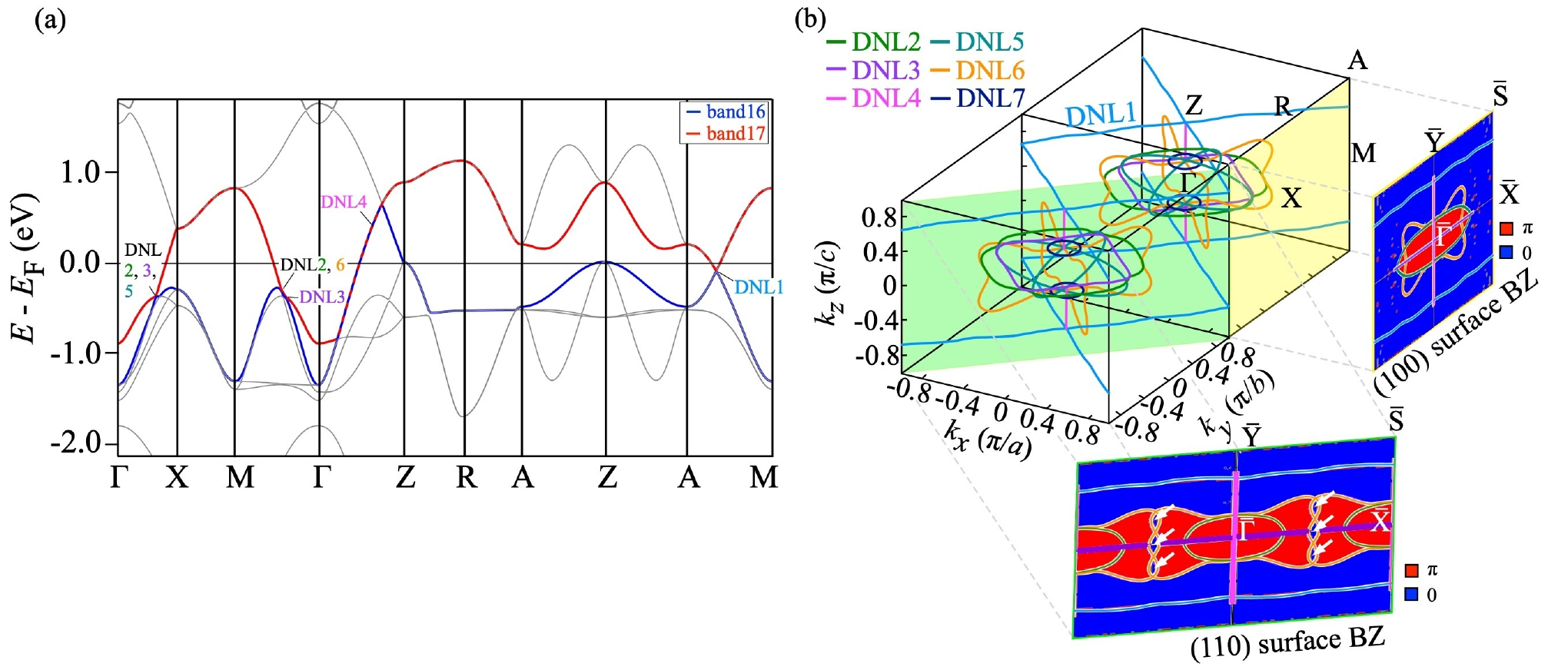}
\caption{(a) Calculated bulk band dispersion along high-symmetry lines in the bulk BZ. Bands 16 and 17 are highlighted in blue and red, respectively. Several DNLs are identified as DNL1 and DNL2--7. (b) DNLs plotted in the 3D bulk BZ. The calculated Zak phases for the (100) and (110) surfaces, together with the projections of the DNLs, are shown in the right and bottom panels, respectively. Area shown by white arrows indicate $\theta = 0$ region surrounded by $\theta = \pi$ region.}
\end{figure}

\clearpage

\begin{figure}[h]
\includegraphics[width=4.8 in]{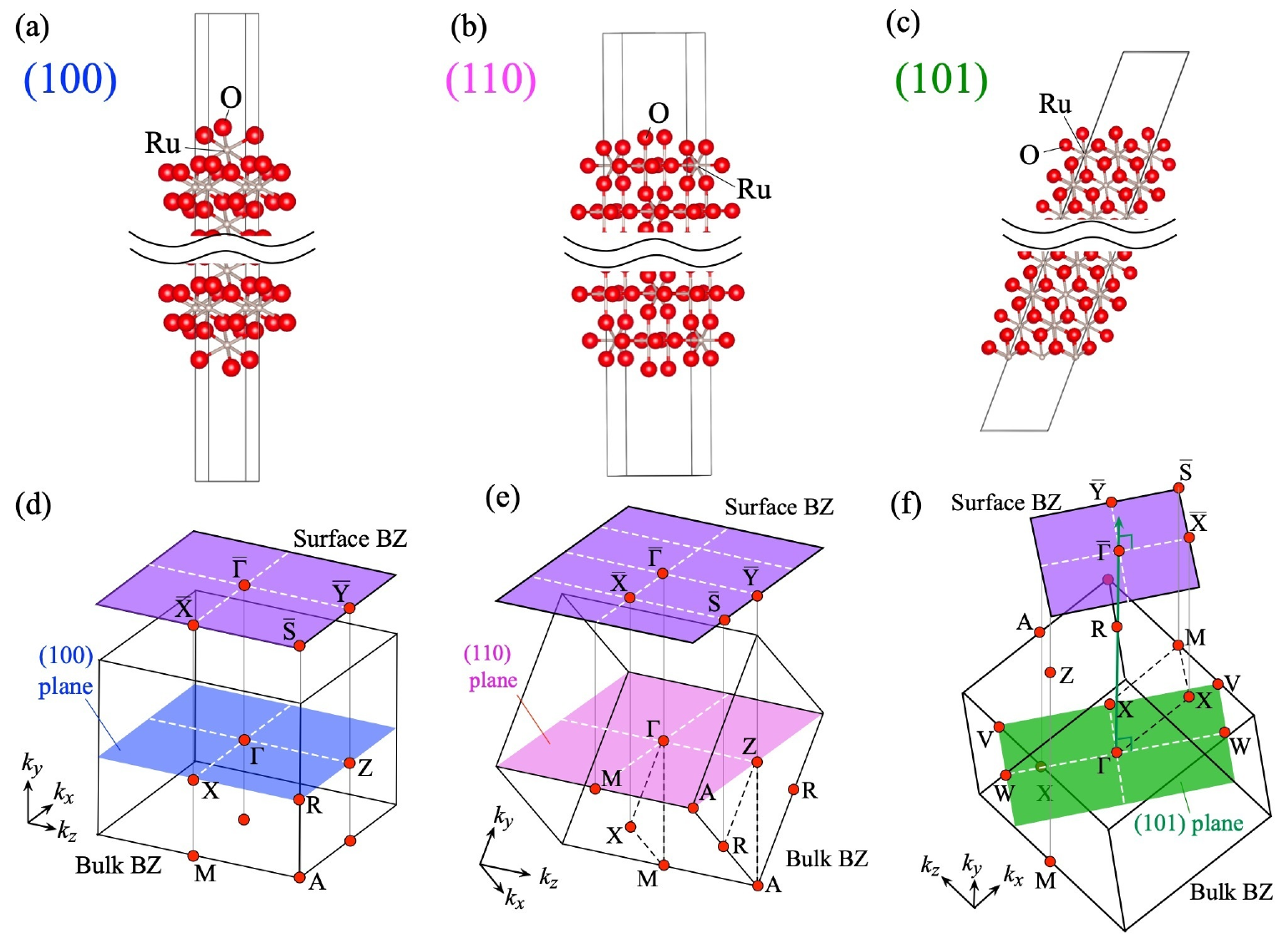}
\caption{\textbf{Slab models for (100), (110), and (101) orientations in RuO$_2$.} (a) Slab structure for the (100), (110), and (101) orientations assumed in the calculation. (c)-(e) Corresponding surface BZ (purple shade) together with the bulk BZ.
}
\end{figure}

\clearpage

\begin{figure}[h]
\includegraphics[width=6.5 in]{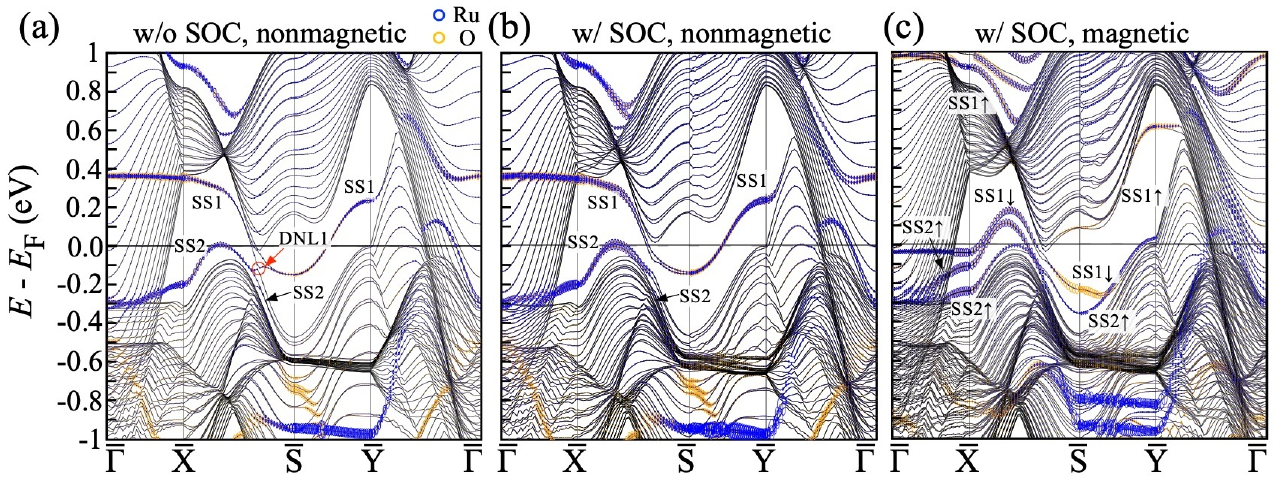}
\caption{(a) Slab calculation without SOC along high-symmetry lines in the surface BZ for the (100) surface. Surface-localized bands are highlighted by colored curves. (b),(c) Same as (a) but including SOC: (b) with and (c) without the constraint of zero total magnetization. In (c), SOC lifts the spin degeneracy of the surface bands via exchange interaction. The notation ``SS1$\uparrow$/$\downarrow$" denotes spin-up and spin-down components of the surface band, SS1.}
\end{figure}

\clearpage

\begin{figure}[h]
\includegraphics[width=5.2 in]{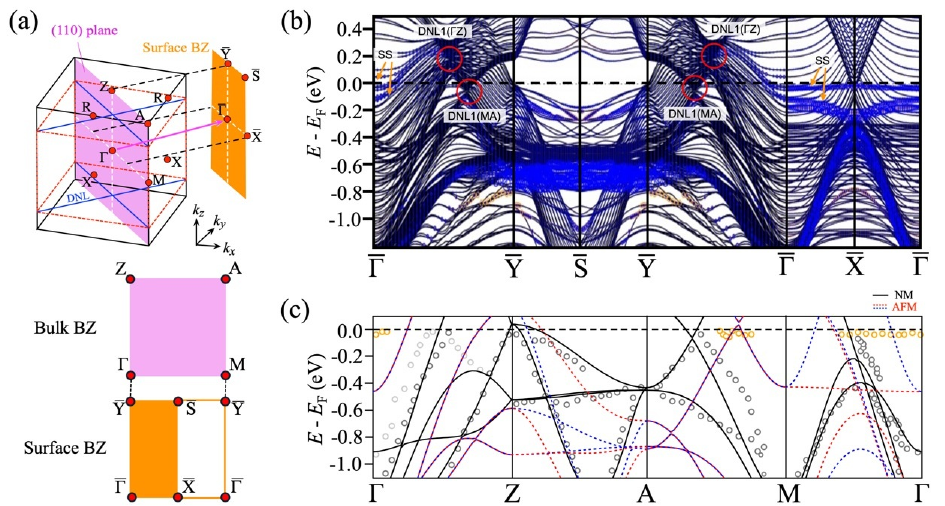}
\caption{(a) Bulk BZ and (110) surface BZ. (b) Slab calculation including SOC, obtained along high-symmetry lines in the surface BZ for the (110) surface. Surface-localized bands are highlighted by colored circles. (c) Same as Fig. 3(e) in the main text. Experimental band dispersion extracted from ARPES along ${\Gamma}\textrm{ZAM}{\Gamma}$ cut. Black and orange circles correspond to bulk and surface bands, respectively. Black solid curves and red/blue dashed curves denote the calculated dispersions for NM and AFM states, respectively.}
\end{figure}

\begin{figure}[h]
\includegraphics[width=5.2 in]{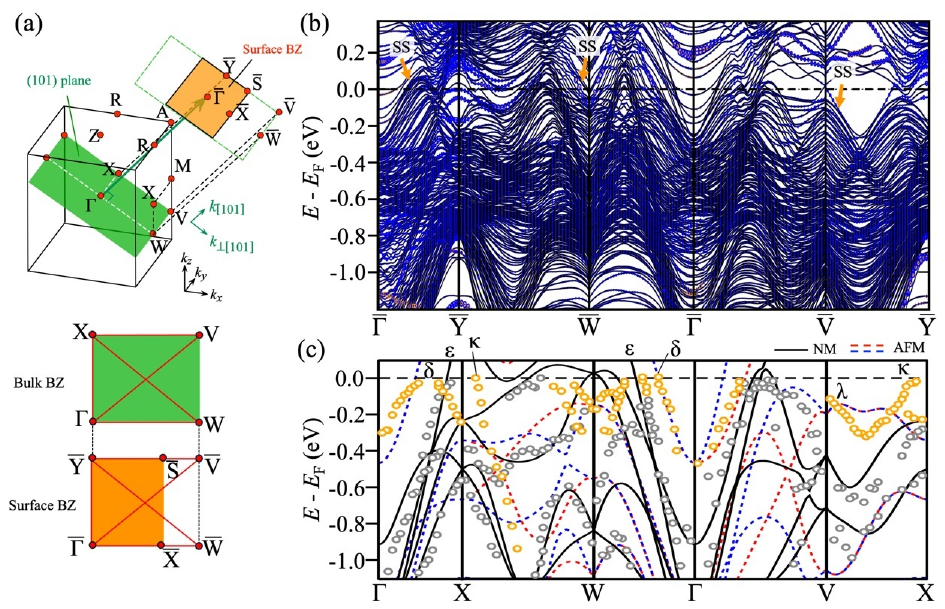}
\caption{(a) Bulk BZ and (101) surface BZ. (b) Same as Fig. 13(b) but for the (101) surface. (c) Same as Fig. 4(e) in the main text.}
\end{figure}

\begin{figure}[h]
\includegraphics[width=4.5 in]{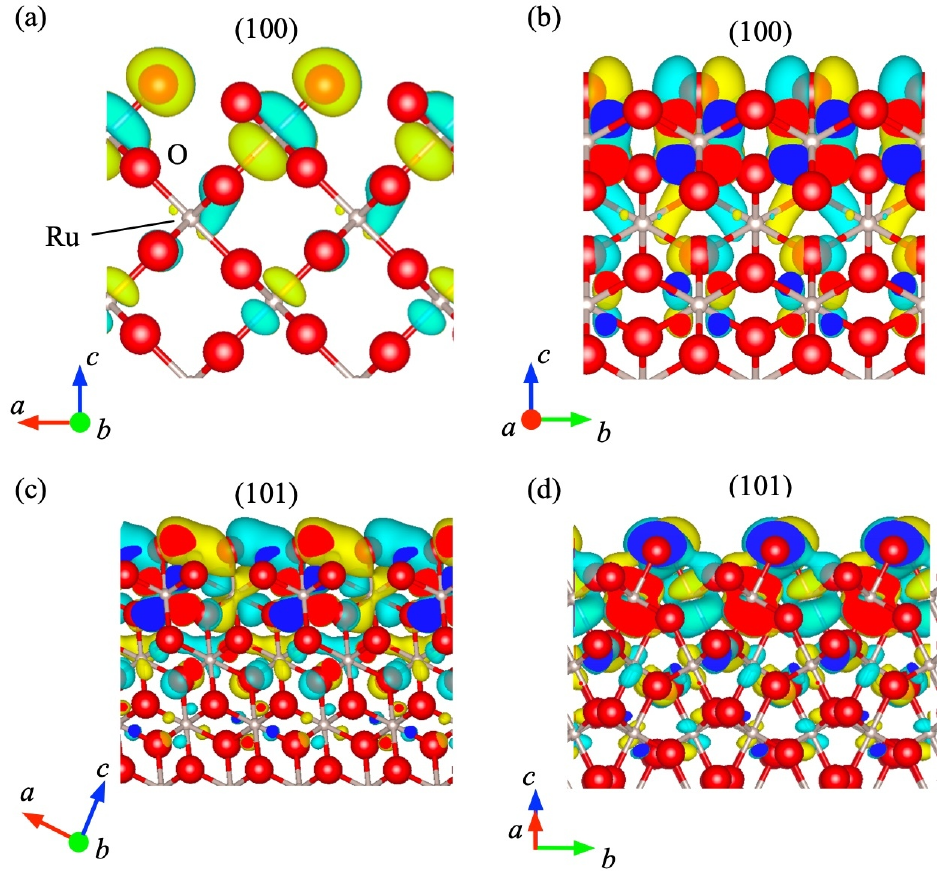}
\caption{(a),(b) Side views of the Wannier orbital of the flat surface state (SS1) from slab calculations including SOC for the (100) surface. Red and white spheres represent O and Ru atoms, respectively. (c) Same as (a) but for a dispersive surface band on the (101) surface.}
\end{figure}

\end{document}